\documentclass[authorversion, nonacm]{acmart}

\usepackage[utf8]{inputenc}

\usepackage[nameinlink,capitalize]{cleveref}

\usepackage{booktabs}
\usepackage{balance}
\usepackage{booktabs}
\usepackage{multirow}
\usepackage{paralist}
\usepackage{array}
\usepackage{bm}
\usepackage{subfig}
\usepackage{graphicx}
\usepackage{makecell}
\graphicspath{{figures/}{plots/}}

\newcolumntype{L}[1]{>{\raggedright\let\newline\\\arraybackslash\hspace{0pt}}m{#1}}
\newcolumntype{C}[1]{>{\centering\let\newline\\\arraybackslash\hspace{0pt}}m{#1}}
\newcolumntype{R}[1]{>{\raggedleft\let\newline\\\arraybackslash\hspace{0pt}}m{#1}}

\newcommand{\statp}[1]{\ensuremath{\left(#1\right)}}
\newcommand{\statreport}[1]{\ensuremath{\left[#1\right]}}

\setcopyright{none}

\acmJournal{TOIS}
\acmVolume{0}
\acmNumber{0}
\acmYear{0000}
\acmArticle{0}
\acmMonth{0}

\begin{document}
\title[A PPA Auction Scheme Using Mouse Cursor Information]{A Price-Per-Attention Auction Scheme Using Mouse Cursor Information}
\titlenote{\textbf{ACM Trans. Inf. Syst. 38, 2, Article 13 (January 2020), 30 pages.}
  DOI: \href{https://doi.org/10.1145/3374210}{10.1145/3374210}}

\author{Ioannis Arapakis}
\affiliation{
 \institution{Telefonica I+D}
 \streetaddress{Plaza de Ernest Lluch i Mart\'{i}n, 5}
 \city{Barcelona}
 \postcode{08019}
 \country{Spain}
}
\email{ioannis.arapakis@tid.es}

\author{Antonio Penta}
\affiliation{
 \institution{ICREA, Universitat Pompeu Fabra and Barcelona GSE}
 \streetaddress{Ramon Trias Fargas, 25-27}
 \city{Barcelona}
 \postcode{08005}
 \country{Spain}
}
\affiliation{
 \institution{TSE}
 \city{Toulouse}
 \country{France}
}
\email{antonio.penta@upf.edu}

\author{Hideo Joho}
\affiliation{
 \institution{University of Tsukuba}
 \streetaddress{1-2 Kasuga}
 \city{Ibaraki}
 \postcode{305-8550}
 \country{Japan}
}
\email{hideo@slis.tsukuba.ac.jp}

\author{Luis A. Leiva}
\thanks{Work partially done while L.A. Leiva was affiliated with Sciling, SL}
\affiliation{
 \institution{Aalto University}
 \postcode{PO BOX 11000}
 \city{Espoo}
 \country{Finland}
}
\email{firstname.lastname@aalto.fi}

\renewcommand{\shortauthors}{I. Arapakis et al.}

\begin{abstract}
Payments in online ad auctions are typically derived from click-through rates, so that advertisers do not pay for ineffective ads. But advertisers often care about more than just clicks. That is, for example, if they aim to raise brand awareness or visibility. There is thus an opportunity to devise a more effective ad pricing paradigm, in which ads are paid only if they are actually noticed. This article contributes a novel auction format based on a pay-per-attention (PPA) scheme. We show that the PPA auction inherits the desirable properties (strategy-proofness and efficiency) as its pay-per-impression and pay-per-click counterparts, and that it also compares favourably in terms of revenues. To make the PPA format feasible, we also contribute a scalable diagnostic technology to predict user attention to ads in sponsored search using raw mouse cursor coordinates only, regardless of the page content and structure. We use the user attention predictions in numerical simulations to evaluate the PPA auction scheme. Our results show that, in relevant economic settings, the PPA revenues would be strictly higher than the existing auction payment schemes.
\end{abstract}


\keywords{Sponsored search; Online advertising; Auctions; Mouse cursor analysis; Direct Displays; User attention}

\maketitle

\section{Introduction}
\label{sec:introduction}

The majority of online advertisements are sold through auctions.
Online ad auctions differ in their baseline format,\footnote{This is especially true when multiple slots are sold at the same time. Google, for example, introduced and still uses the celebrated Generalized Second Price (GSP) auction format,
whereas Facebook follows the classic Vickrey-Clark-Groves (VCG) format.
For single-slot ad sales, both GSP and VCG are identical.}
but all existing formats adopt either
\emph{pay-per-impression} (PPI) or \emph{pay-per-click} (PPC) schemes.
While important parts of the market still adopt the PPI, in recent
years the market has increasingly shifted toward the PPC scheme, which
presents several advantages for advertisers.\footnote{The PPC format is predominant for ads which are sold through auctions.
With posted prices, PPI systems tend to be preferred instead; see, e.g., \citet{ChoiSurvey2018}.} First, it insures against the risk of paying for ineffective ads, as under the PPC scheme advertisers pay only if their ads are effectively clicked.\footnote{Google reported that more than half of the ad impressions go unnoticed by users \cite{google2014}.} Second, it provides a better alignment between the platform's incentives and the advertisers' objectives, since it links the revenue of the former directly to the Click-Through-Rate (CTR), which represents the main performance measure targeted by advertisers. Perhaps for these reasons, PPC systems also tend to induce lower costs per click than PPI on average.

CTRs, however, are often an imperfect measure of ad performance. If clicks are one of the main objectives of the advertisers, they need not be the only, nor the most important.\footnote{\citet{blake2015}, for example, discuss evidence which suggests that advertisers are willing to pay to post an ad beyond the clicks it is likely to generate.} This is especially the case for advertisers whose campaigns aim to generate mainly brand or product awareness, rather than to induce direct online sales --- which is the case for many, if not most, of the highest-value advertisers. In these cases, advertisers ultimately care more about making sure that consumers notice the ad, and may thus attach a value to grabbing the consumer attention beyond the click it may or may not trigger. Therefore, if consumer attention to ads could be measured in an accurate and reliable way, not only would it provide a more effective target for advertiser campaigns, but it could also serve as a basis to define auctions which directly price user attention, thereby further aligning the platforms' incentives with the objectives of the advertisers. This creates an opportunity for a novel ad pricing paradigm to take root; one in which ads are paid \emph{only if they are actually noticed}.

In this article, we introduce a novel auction format based on a \emph{pay-per-attention} (PPA) scheme, in which advertisers' payments is proportional to the probability that their ad is noticed by users. To make this PPA format possible, we also contribute a scalable diagnostic technology to measure users' attention to ads, using client-side user interactions derived from mouse cursor movements (without such a diagnostic technology, the PPA auction format would not be feasible in practice). Finally, we show both analytically and through numerical simulations that the two innovations combined may enable new ad platforms to extract revenues which are strictly higher, and in any environment never lower, than the existing PPI and PPC auction payment schemes.

We begin by introducing our novel PPA auction scheme, a second-price auction which takes measures of users' attention as input, similar to the way existing PPC formats take clicks as input.
We show that, like its PPI and PPC counterparts, the PPA auction is \emph{strategy-proof} (that is, bidding one's own value is optimal regardless of the bids placed by others) and \emph{efficient} (that is, the ad slot is always allocated to the advertiser with the highest valuation), and that it has desirable revenue properties. In particular, we show that in relevant economic environments the PPA revenues are strictly larger, and in any environment never lower, than those of the PPI and PPC formats. Environments in which the PPA yields strictly higher revenues than the PPI or PPC formats include, for example, environments in which advertisers' valuations are positively correlated with the probability of their ads being noticed, or if at least some of the advertisers are subject to framing effects, in the sense that they bid only taking into account the components of the value which are made salient by the different auction formats.

Then, we introduce our diagnostic technology, which is specifically designed to measure consumer attention to ads in sponsored search. More concretely, it is based on mouse cursor movements analysis and builds upon previous work examining user engagement with direct displays in web search~\cite{Arapakis:2016:PUE:2911451.2911505}.
We conduct a crowdsourced user study and collect mouse cursor data from participants who interacted with instrumented Search Engine Result Pages (SERPs) in brief transactional search tasks with Google Search. The SERPs contained, among other elements, sponsored ads that were served under different formats and positions.
We use the collected data to train several baseline machine learning models and a recurrent neural network model to predict user attention to SERP ads. We further demonstrate noticeable improvements by our recurrent neural network (which uses raw mouse cursor data) over the baseline machine learning models (which rely on ad-hoc and domain-specific features).

Next, we provide numerical simulations based on a variation of the statistical model proposed by \citet{AusubelBaranov2018}, both to exemplify the main insights highlighted by our theoretical findings and to illustrate their significance in the context of the distribution of attention probabilities observed in our online user study.
Taken together, our results show that, combined with an effective and reliable technology to predict users' attention, such as the one proposed in this work, the PPA auction may draw from sources of economic surplus which may be precluded to the existing formats. Our analysis focuses on some of such possibilities, but the main innovations of our work open other directions for future research, which we discuss in the final section of this article.

\section{Related Work}
\label{sec:related_work}

In what follows, we provide a review of the state of the art in the two areas which our work bridges:
online ad auctions and mouse cursor analysis.

\subsection{Online Ad Auctions}
\label{ssec:RL_auctions}

A large body of work has analysed the various auction formats used to sell online ads.
For the sake of conciseness, we only review work that focuses on the two most common formats,
namely the Generalized Second Price (GSP) auction (used e.g., by Google and Taobao)
and the Vickrey-Clarke-Groves (VCG) mechanism (used e.g., by Facebook and Quora).
For the case of a single slot, both formats coincide with the sealed-bid second-price auction.

\subsubsection{Models with Competitive Bidding}
The VCG is a classic and well-known mechanism in the economics literature,
introduced by \citet{vcg1}, \citet{vcg2} and \citet{vcg3}.
It is both strategy-proof (that is, bidding one's own value is optimal regardless of the bids placed by others)
and efficient (that is, the ad slot is always allocated to the advertiser with the highest valuation).
The study of the GSP was pioneered by \citet{aggarwal2006}, \citet{EOS2007} and \citet{varian2007},
who introduced an equilibrium refinement of the GSP
(which induces the same revenues and allocations as the VCG),
and further refinements were later provided by others~\cite{edelman2010, milgrom2018}. Interestingly, most of the literature on the GSP auctions studies environments with complete information.
A notable exception considered a standard independent private values environment~\cite{gomes2014}.
\citet{borgers2013} also maintain the complete information assumption,
but consider a more general model of CTRs and valuations.
Without resorting to \citet{EOS2007} and \citet{varian2007}'s refinement,
it provides a more critical view of the GSP; see also \citet{PaesLeme2010}.
\citet{athey2012} introduced uncertainty over quality scores in a model with competitive bids,
to account for the fact that existing models assume that bids are customised for a single user query
but in practice queries arrive more quickly than advertisers can change their bids.

\subsubsection{Coordinated and Collusive Bidding}
Recent literature studied online ad auctions with coordinated or collusive bidding,
to account for the increasing diffusion of digital marketing agencies and of agency trading desks.
\citet{mansour2012}, for example, pointed at the potential risk of collusive bidding
that intermediaries pose for online ad auctions, and focused on the ad exchanges used for sponsored ads.
\citet{DGP2018} studied agency bidding in both the GSP and VCG auctions,
allowing for the co-presence of collusive and competitive bidders.
The earlier literature on ``bidding rings'' mostly focused on single-unit mechanisms
in which either non-cooperative behavior is straightforward~\cite{Mailath1991},
or assuming that the coalition includes all bidders~\cite{McAfee1992, Hendricks2008}.
In multi-unit settings, \citet{bachrach2010} studied collusive bidding in the VCG,
but from a cooperative game theory perspective.

\subsubsection{Variations of the Baseline Formats}
In the baseline GSP auction, Yahoo! initially ranked advertisers by bids.
Then Google adopted a ranking based on value per impression,
in which bids are weighted by quality scores, designed to increase revenues.
Other variations of the baseline GSP apply instead reservation prices in the spirit of \citet{Myerson1981},
thereby offering a compromise between efficiency and revenue maximization.
\citet{ostrovsky2011} studied the effects of applying optimal per-impression reserve price
and \citet{Roberts2013} studied the revenue optimal auction,
empirically showing that it led to good trade-offs between revenue and other objectives.
Theoretical results on optimal trade-offs are also available~\cite{bachrach2014, JehielAndLamy2015}.
Finally, \citet{Thompson2013} studied a variety of ways of increasing revenue,
including optimal reserve prices as well as alternative ranking algorithms,
and others have studied the welfare effects of reservation prices in various settings~\cite{edelman2010, LahaieMcAfee2011, AtheyEllison2011}.

\subsubsection{PPC, PPI, and Consumer Attention}
A few works compare PPC and PPI schemes,
but mostly in non-auction pricing models of ads --- where, unlike auction settings,
PPI schemes are predominant; see \citet{ChoiSurvey2018} and references therein.
In deterministic settings, \citet{Mangani2004} compares revenues between PPC and PPI schemes,
and \citet{Fjell2009} determines the optimal choice between both schemes,
which have been further studied with stochastic arrivals of viewers and advertisers~\cite{Najafi2014, Fridgeirsdottir2018}.

\subsection{Mouse Cursor Analysis}
\label{ssec:mouse_cursor_analysis}

The construct of attention, broadly indicating a high degree of involvement in a given activity, has become a common currency on the Web nowadays. Objective measurements of attentional processes are increasingly sought after by both the media industry and scholar communities to explain or predict user behavior. In recent years, a large body of research~\cite{Shapira:2006:SUK:1141277.1141542, Guo:2008:EMM:1390334.1390462, Guo:2010:RBJ:1835449.1835473, Guo:2012:PWS:2396761.2398570, Huang:2012:USU:2207676.2208591, Navalpakkam:2013:MME:2488388.2488471, Lagun:2014:DCM:2556195.2556265, Liu:2015:DUD:2766462.2767721, MARTINALBO2016989, 8010344} has demonstrated the utility of mouse cursor analysis as a low-cost and scalable proxy of visual attention. In line with this evidence, several works have investigated closely the user interactions that stem from mouse cursor data for various use cases, such as web search~\cite{Guo:2008:EMM:1390334.1390462, Guo:2010:RBJ:1835449.1835473, Guo:2012:PWS:2396761.2398570, Lagun:2014:DCM:2556195.2556265, Liu:2015:DUD:2766462.2767721, Arapakis:2016:PUE:2911451.2911505, 8010344} or web page usability evaluation~\cite{Atterer:2006:KUM:1135777.1135811, Arroyo:2006:CPE:1125451.1125529, Leiva:2011:RWD:2037373.2037467}. In what follows, we review those research efforts that have focused on mouse cursor analysis to predict user interest and attention. We deliberately leave out works that investigate ad impression forecasting using click logs or query traffic features~\cite{Kolesnikov:2012:PCN:2396761.2398688, Zilong2016, Nath:2013:AIF:2488388.2488470, Guo:2010:RBJ:1835449.1835473, Lagun:2014:TBM:2600428.2609631, Zhai:2016:DLA:2939672.2939759, Mao:2018:CCM:3209978.3210060}, since our approach relies solely on implicit, online interaction signals instead of historical, click-through data. Furthermore, while the PPA scheme we propose is independent of how user attention is detected, the diagnostic technology we introduce here addresses a desktop setting. Hence, works on ad noticeability or attention in mobile browsing~\cite{Barbieri:2016:IPU:2872427.2883092, Li:2017:TMI:3041021.3054182, Wang:2018:LSM:3209978.3210099, Mao:2018:TMU:3176349.3176891, Grusky:2017:MSA:3025453.3025916, Lagun:2016:UMS:2983323.2983853} also fall outside the scope of this article.

\subsubsection{Measuring User Interest}
\label{ssec:interest}

For a long time, user models of scanning behaviour in SERPs have been assumed to be linear, as users tend to explore the list of search results from top to bottom. Today this is no longer the case, since SERPs now include several heterogeneous modules (direct displays) such as image and video search results, featured snippets, or ads~\cite{Arapakis15_cikm}.
To account for this SERP heterogenity, \citet{Diaz13} proposed a generalization of the classic linear scanning model which incorporated ancillary page modules. Here, a user interaction log is represented by a sequence of visited (mouse-hovered) SERP modules. This model can help improve SERP design by anticipating searchers' engagement patterns given a proposed arrangement of the SERP. However, this model is not designed to effectively measure if a user is actually paying attention to ads and does not exploit the potential information encoded in mouse coordinates.

Early research considered simple, coarse-grained features derived from mouse cursor data to be surrogate measurements of user interest, such as the amount of mouse cursor movements~\cite{Shapira:2006:SUK:1141277.1141542} or the mouse cursor's ``travel time''~\cite{Claypool:2001:III:359784.359836}. More recent work has adopted fine-grained mouse cursor features, which have been shown to be more effective. For example,~\citet{Guo:2008:EMM:1390334.1390462} found differences in mouse cursor distances between informational and navigational queries, and could classify the query type using mouse cursor movements more accurately than using clicks alone. In a similar vein,~\citet{Guo:2010:RBJ:1835449.1835473} examined implicit interaction signals like mouse cursor movements, hovers, and scrolling activity to accurately infer search intent and interest in SERPs. They focused on automatically identifying a user's research or purchase intent based on features of the interaction.
These approaches have been directed at predicting general-purpose web search tasks like search success~\cite{Guo:2012:PWS:2396761.2398570} or search satisfaction~\cite{Liu:2015:DUD:2766462.2767721} and, in that respect, lack the granularity in predicting attention with particular direct displays of a SERP, such as ads, that our proposed modelling approach achieves.

In a more recent work,
\citet{Huang:2012:ISM:2348283.2348313} and~\citet{Speicher:2013:TPR:2505515.2505703} modelled mouse cursor interactions on SERPs, by extending click models to compute more accurate relevance judgements for the search results.
In a similar line, \citet{Huang:2011:NCN:1978942.1979125} sought to understand results relevance and search abandonment
by mining mouse cursor behaviour on SERPs.
They showed that the mouse cursor position is mostly aligned to eye gaze, especially on SERPs,
and that could be used as a good proxy for predicting good and bad abandonment.
\citet{Diriye:2012:LSS:2396761.2398399} extended this work and investigated the effectiveness of mouse cursor interactions for predicting the reasons for observed search abandonment, whether it was because the user's information need was satisfied or because they were dissatisfied with the search results.
\citet{Feild:2010:PSF:1835449.1835458} considered mouse movements, among other sensory and log-based features, to predict success and frustration in an information seeking task and \citet{Guo:2012:PWS:2396761.2398570} examined mouse cursor movements to identify patterns of examination and interaction behaviour that indicated search success.
Finally, \citet{Guo:2012:BDT:2187836.2187914} looked at mouse cursor interactions after the click onto the landing page and found that these post-click interactions (e.g., mouse cursor movements, dwell time) correlate with document relevance. They showed that a post-click behaviour model is more effective than simply using dwell time for computing document relevance scores.

\subsubsection{Measuring User Attention}
\label{ssec:attention}

Most research studies assume that eye fixation means examination, including studies from industry~\cite{Brightfish18}.
However, \citet{Liu:2014:SRT:2661829.2661907} noticed that almost half of the search results fixated by users are not being read, since there is often a preceding skimming step
in which the user quickly looks at the search result without reading it.
Based on this observation, they propose a two-stage examination model:
a first ``from skimming to reading'' stage and a second ``from reading to clicking'' stage.
Interestingly, they showed that both stages can be predicted with mouse movement behaviour,
which can be collected at large scale.

Cursor movements can therefore be used to estimate user attention on SERP components,
including traditional snippets, aggregated results, maps, and advertisements, among others.
However, works that employ mouse cursor information to predict user attention with specific elements within a web page have been scarce. This is understandable, considering the inherently complex nature of the mouse cursor data and the difficulty in constructing ground-truth labels at scale. Despite these challenges, some of the early work by~\citet{Arapakis:2014:UEO:3151365.3151368, Arapakis:2014:UWE:2661829.2661909} investigated the utility of mouse movement patterns to measure within-content engagement on news pages and predict reading experiences. \citet{Lagun:2014:DCM:2556195.2556265} introduced the concept of frequent cursor subsequences (namely \emph{motifs}) in the estimation of result relevance. Although their work proposes a more general approach to mouse cursor pattern analysis, it does not target specific user engagement proxies such as attention. Similarly,~\citet{Liu:2015:DUD:2766462.2767721} applied the motifs concept to SERPs in order to predict search result utility, searcher effort, and satisfaction at a search task level. Their approach assumes a uniform engagement with all parts of the page and, in that sense, lacks the desired granularity in the analysis of mouse cursor interactions.

To our knowledge, the closest work to ours is that of~\citet{Arapakis:2016:PUE:2911451.2911505}, which investigated user engagement with direct displays on SERPs, and more specifically with the Knowledge Graph display.\footnote{The Knowledge Graph is a card-like direct display that appears, for some informational queries, at the top-right part of the SERPs, comprising information related to the named entities related to the current user query.}
Similarly, we implement a predictive modelling framework to measure user attention to ads, which can be seen as a particular instance of direct displays. However, our work differs significantly from theirs in three key respects. First, we compare their machine learning model (which relies on ad-hoc and domain-specific features)
against a recurrent neural network model (which uses raw mouse cursor data) to predict user attention, and demonstrate noticeable improvements by our recurrent neural network model. Second, we examine the performance of our predictive modelling approach w.r.t. sponsored ads served under (i)~different formats and (ii)~different positions within a SERP and, thus, significantly expand on the original application and findings by~\citet{Arapakis:2016:PUE:2911451.2911505}. Last, we examine interaction effects in performance between our predictive model and different demographic attributes, such as gender and age, which highlights new opportunities for market segmentation.

\subsection{Summary}

Overall, a few works have studied the relative performance of PPI and PPC auction schemes, but we are not aware of theoretical work which has provided explanations of the difference in performance which seem to emerge from the data, nor of why most of the auction platforms tend to favour a PPC over a PPI scheme (as explained above, the opposite is true in non-auction pricing settings). To the best of our knowledge, existing theoretical analysis of online auctions assume that bidders attach no value to obtaining an ad slot unless it is clicked. This assumption has proven to be a good proxy for the theoretical questions addressed by the literature, but it overlooks aspects of bidders' valuations which are important in many markets, where the ads value is more directly related to their ability to attract users' attention, beyond their clicks. There is thus an opportunity to devise a more effective ad pricing paradigm, in which ads are paid only if they are actually noticed. However, for our novel PPA auction scheme to become feasible, a scalable diagnostic technology that estimates user attention to ads becomes necessary. We make it possible with a recurrent neural network model that relies exclusively on implicit interaction signals derived from mouse cursor movements, and show noticeable improvements over the state of the art. We also show that our model generalizes to different ad formats and different positions within a SERP.

\section{The PPA Auction Scheme}
\label{sec:ppa_auction}

In line with the motivation discussed in the \nameref{sec:introduction} section, the auction model we introduce in this section explicitly accounts for the possibility that advertisers value an ad slot's ability to grab the user attention, beyond the clicks that it may generate~\cite{blake2015}. We show that the PPA auction is both efficient (that is, the slot is always allocated to advertiser with the highest valuation) and strategy-proof (that is, bidding one's own value is optimal regardless of the bids placed by others). We also compare the revenue properties of the PPA auction with those of its standard PPI and PPC counterparts, under various settings. Analytic results show that PPA outperforms its PPI and PPC counterparts both when bidders' valuations are positively correlated with their ability to attract users' attention, and in the presence of \emph{framing effects}, under which bidding strategies are assumed to affect advertisers' bidding strategies, through the component of their valuations which are made salient. Both of these features (i.e., positive correlation and framing effects) are expected to be present in most relevant economic settings. Moreover, our results also show that in environments in which these conditions are not met, revenues under the PPA auction are in any case no lower than under the standard PPI and PPC formats.

\subsection{Auction Environment}
\label{ssec:environment}
For simplicity, we focus on the case of a single slot for sale (extensions to the multiple-slot case are the subject of future work). We thus consider a setting in which a single ad slot needs to be allocated to one of $n \geq 2$ bidders, indexed by $i \in I = \left\{1, \dots, n\right\}$. We let $p$ denote the probability that a particular consumer visiting a web page notices the posted ad, and let $x$ denote its click-through rate (CTR), conditional on being noticed (the case of bidder-specific attention probabilities is discussed below).\footnote{In most models, CTRs are typically expressed unconditional on the ad being noticed. The standard \emph{unconditional} CTR in this model is thus equal to $\text{CTR} = p x$.} For each $i \in I$, let $\gamma_{i} \geq 0$ denote the probability that $i$ realizes a sale conditional on the consumer having clicked on the ad, and let $q_{i}$ denote the probability that $i$ realizes a sale conditional on the ad being noticed but not clicked. Finally, we let $v_{i}$ denote the value of a sale for bidder $i$.

We assume that both $x$ and $p$ are commonly known by advertisers, whereas $\gamma_{i}$, $q_{i}$ and $v_{i}$ are $i$'s private information, which follow mutually independent distributions $\Gamma, Q$, and $F$, respectively, i.i.d. across bidders. Bidders' types are thus three-dimensional, $t_{i} = \left(\gamma_{i}, q_{i}, v_{i}\right)$
. The total unconditional expected value of obtaining the slot for bidder $i$ with type $t_{i}$
therefore is
\begin{equation}\label{eq:EV}
\text{EV}_{i} = p \left[x\gamma_{i} + \left(1 - x\right) q_{i}\right] v_{i}.
\end{equation}

The most standard auction format in this environment is the second-price auction:
bidders submit bids, denoted by $b_{i} \geq 0$; the highest bidder gets the slot and pays the second-highest bid, with ties broken randomly with a fair lottery over the highest bidders.
Existing formats differ in the payment rule: they are typically based on either a PPC or PPI scheme.
In the PPC scheme, a bidder pays only if their ads are clicked;
in a PPI scheme, the bidder pays for the very fact of obtaining the ad slot.

In the PPA scheme we propose, a bidder's payment is proportional to the probability their ad is noticed. Clearly, the possibility of defining such a payment rule presumes that the attention probability $p$ is observable. For this reason, the next two sections will be dedicated to the development of a diagnostic technology to detect attention probabilities. Here, however, we first focus on the properties of the PPA auction, under the assumption that attention probabilities are observable. We will return to the matter of how attention probabilities can be predicted in Sections~\ref{sec:user_study} and \ref{sec:predicting_attention}, and then will combine our theoretical and experimental results with numerical simulations in \autoref{sssec:numerical_simulations}.

Formally, letting $b_{-i}^{\max}$ denote the highest opponent's bid, i.e. $b_{-i}^{\max}
:=\max_{j\neq i}\left\{b_{j}\right\}$,
in the PPA second-price auction bidder $i$'s (expected) payoff,
conditional on a particular profile of bids $b=\left(b_{i},b_{-i}\right)
\in R^{n}$ and $p$, is:

\begin{equation}\label{eq:PPA}
u_{i}^{\text{PPA}}\left(b\right) = \left\{
\begin{array}[c]{lr}
\frac{\text{EV}_{i}
-pb_{-i}^{\max}}{\#\left\{ j:b_{j}=b_{i}\right\}}
  & \text{if}~b_{i} \geq b_{-i}^{\max}\\
0 & \text{if}~b_{i} < b_{-i}^{\max}
\end{array}
\right.
\end{equation}

For later reference, we contrast this with the payoff functions which result from the PPI and PPC schemes, respectively:

\begin{equation}\label{eq:PPI}
u_{i}^{\text{PPI}}\left(b\right) = \left\{
\begin{array}[c]{lr}
\frac{\text{EV}_{i}
-b_{-i}^{\max}}{\#\left\{ j:b_{j}=b_{i}\right\}}
  & \text{if}~b_{i} \geq b_{-i}^{\max}\\
0 & \text{if}~b_{i} < b_{-i}^{\max}
\end{array}
\right.
\text{, \quad and \quad}
u_{i}^{\text{PPC}}\left(b\right) = \left\{
\begin{array}[c]{lr}
\frac{\text{EV}_{i}
-pxb_{-i}^{\max}}{\#\left\{ j:b_{j}=b_{i}\right\}}
  & \text{if}~b_{i} \geq b_{-i}^{\max}\\
0 & \text{if}~b_{i} < b_{-i}^{\max}
\end{array}
\right.\text{.}
\end{equation}
Note that the term in the denominators in equations \eqref{eq:PPA} and \eqref{eq:PPI} is equal to the number of bidders who are placing the highest bid, as $b_i \geq b^{max}_{-i}$ in that case.

Existing analysis of the PPI and PPC formats are special cases of these payoff functions by setting $q_{i}=0$ in \cref{eq:EV}. As discussed in the previous footnote, in most models of PPC auctions it is standard to work with the \emph{unconditional} click-through rate, or $\text{CTR} = p x$, and valuations are expressed in terms of \emph{value-per-click} (VPC), or $\text{VPC}_{i}=\gamma_{i}v_{i}$ in our setting; see e.g., \citet{varian2007,EOS2007}. Models of PPI-schemes instead typically ignore specifying the various components of the overall value (namely CTRs, attention probabilities, and value per click), which are irrelevant in those settings, and work directly with the \emph{value-per-impression} (VPI), which in our setting corresponds to $\text{VPI}_{i}=\text{EV}_{i}$. In our setting, it is also convenient to define the \emph{valuation per attention} (VPA) for each $i$, as $\text{VPA}_{i}:=\left[x \gamma_{i} + \left(1 - x\right) q_{i}\right] v_{i}$.

\subsection{Main Properties}
\label{ssec:main_properties}

In what follows we show that the PPA second-price auction thus defined shares the main properties of its PPI and PPC counterparts, namely strategy-proofness and efficiency. As standard in the literature, \emph{strategy-proofness} refers to the property that bidding one's own valuation is a dominant strategy, i.e. it is optimal independent of bids placed by others; whereas \emph{efficiency} instead refers to the standard ex-post concept, which ensures that the slot is always assigned to the highest-valuation advertiser (both per-attention, and overall).

\begin{theorem}\label{thm:PPA}
In the PPA second-price auction, bidding $b_{i}^{\text{PPA}} = \text{VPA}_{i}$ is a dominant strategy for every player. It follows that, in this equilibrium, for each realisation of types $\left(\gamma_{i}, q_{i}, v_{i}\right)_{i \in I}$ the slot is assigned to the advertiser with the highest valuation per attention (both per-attention, and overall).

 \begin{proof}
We first show that $b_{i}^{\text{PPA}} = \text{VPA}_{i}$
dominates any bid $b_{i} > \text{VPA}_{i}$: if $b_{-i}^{\max} > b_{i} > \text{VPA}_{i}$, both
$b_{i}^{\text{PPA}}$ and $b_{i} > \text{VPA}_{i}$ yield a payoff of zero, since the slot is
allocated to a different bidder; if $b_{-i}^{\max} \leq \text{VPA}_{i}$, both
$b_{i}^{\text{PPA}}$ and $b_{i} > \text{VPA}_{i}$ suffice to win the auction, yielding a
payoff of $p\left[\text{VPA}_{i} - b_{-i}^{\max}\right] \geq 0$; if $b_{-i}^{\max}
\in (\text{VPA}_{i}, b_{i}]$, then $b_{i}^{\text{PPA}}$ loses the auction and yields a payoff
of $0$, while $b_{i}$ wins the auction yielding a payoff of $p\left[
\text{VPA}_{i} - b_{-i}^{\max}\right] < 0$. Hence, $b_{i}^{\text{PPA}} = \text{VPA}_{i}$ dominates any
bid $b_{i} > \text{VPA}_{i}$. A similar argument shows that it also dominates any
$b_{i} < \text{VPA}_{i}$. Thus, if everybody follows the dominant strategy $b_{i}
^{\text{PPA}} = \text{VPA}_{i}$, the rules of the auction imply that the winner is in the set
$\arg\max_{i \in I} \text{VPA}_{i}$.
\end{proof}
\end{theorem}

\begin{corollary}
Relabeling players if necessary so that $\text{VPA}_{1} \geq \text{VPA}_{2} \geq \ldots \text{VPA}_{n}$,
for each realisation of types the revenues in this auction are equal to
$p \text{VPA}_{2}$, and bidders' payoffs are equal to $p\left[\text{VPA}_{1} - \text{VPA}_{2}
\right]  $ for the highest bidder, and $0$ for all others.
\end{corollary}

\subsection{Revenue Comparisons}
\label{ssec:revenue_comparisons}

We now compare the revenues of the PPA second-price auction with those of its PPI and PPC counterparts, in different settings. We show that, with fully sophisticated bidders (i.e., those who try to bid optimally) both the PPI and PPC auctions induce the same revenues and payoffs for all bidders as the PPA. Such equivalence between PPI and PPC may seem at odds with the industry's consensus that the latter induces on average lower costs-per-click, but it stems from the fact that all parameters and distributions in the model are kept constant across different payment schemes, and perfectly anticipated by a fully sophisticated bidder.

To make sense of the common wisdom, one has to enrich the model so as to take into account the possibility that CTRs and attention probabilities vary systematically across payment schemes, possibly due to the platforms incentives to improve their performance, thereby increasing the overall value for the advertisers. We abstract from these possibilities here, which would provide further reasons to prefer the PPA scheme, and focus instead on two simpler variations of the baseline model; namely heterogeneity in the attention probabilities and the possibility of framing effects associated with the different payment formats.

In this section we offer some analytic results. In \autoref{sssec:numerical_simulations} we will illustrate the significance of these results for relevant distributions of attention probabilities, through some numerical simulations specifically calibrated on the distributions of attention probabilities observed in the online user study from Sections~\ref{sec:user_study} and \ref{sec:predicting_attention}.

\subsubsection{Analytic Results}
\label{sssec:analytic_results}

\paragraph{Fully Sophisticated Bidders}

Existing theoretical analysis of PPI and PPC auctions do not consider the possibility that advertisers may value an ad beyond it being clicked. They thus consider an ex-ante value equal to $\text{CTR} \cdot \text{VPC}_{i}$, where CTR is the unconditional click-through rate, and $\text{VPC}_{i}$ is the value-per click.\footnote{See e.g.,\citet{varian2007,EOS2007}, which boil down to this in the special case of a single-slot on sale. In terms of our model, as we mentioned, this amounts to assuming $q_{i}=0$ for all $i$, and letting $\text{CTR} = p x$ and $\text{VPC}_{i} = \gamma_{i}v_{i}$.} In such models, it is well-known that bidding $\text{CTR} \cdot \text{VPC}_{i}$ in the PPI, and $\text{VPC}_{i}$ in the PPC are dominant strategies. But if bidders are concerned with their ad being noticed, and if they are fully sophisticated, then their optimal bids would be higher than these, to take into account the possibility that, for some realisation, ads may be noticed but not clicked. The real incentives to win the auction therefore exceeds the value of being clicked, which would be $\text{VPC}_{i}=\gamma_{i}v_{i}$, and a fully sophisticated bidder incorporates
this by raising their  bid by an extra $\left(  1-x\right)  \frac{q_{i}}{x}v_{i}$. This logic is confirmed by the next analytic result:

\begin{proposition}
\label{prop RevEquiv1}
With fully sophisticated bidders, the PPI and PPC
auctions have dominant strategies $b_{i}^{\text{PPI}}=p\left[
x\left(  \gamma_{i}-q_{i}\right)  +q_{i}\right]  v_{i}$ and $b_{i}
^{\text{PPC}}=\left(  \gamma_{i} - q_{i} + \frac{q_{i}}{x}\right)  v_{i}$, respectively.
In the corresponding equilibria, the PPI and PPC induce the same allocation,
revenues and payoffs for every bidder as the PPA auction.

\begin{proof}
If bidders fully realize the payoff implications of the auction rule, their
payoffs in the PPI and PPC auctions when $b_{i}>b_{-i}^{\max}$ are equal to
$u_{i}^{\text{PPI}}\left(\text{win}\right) = p \text{VPA}_{i}-b_{-i}^{\max}$ and
$u_{i}^{\text{PPC}}\left(\text{win}\right) = p x\left(\frac{\text{VPA}_{i}}{x}-b_{-i}^{\max}\right)$, respectively.
For both auctions, payoffs are $u_{i}^{\text{PPC}}\left(\text{win}\right)  /\#\left\{  j:b_{j}=b_{-i}^{\ast}\right\}$
if $b_{i}=b_{-i}^{\max}$, and $0$ if $b_{i}<b_{-i}^{\max}$.
By the same logic as the proof of \autoref{thm:PPA},
it follows that $b_{i}^{\text{PPI}} = p \text{VPA}_{i}$ and $b_{i}^{\text{PPC}} = \frac{\text{VPA}_{i}}{x}$.
Hence, for each realisation of types, revenues are equal to the second highest
bid in the PPI, and by that times the unconditional CTR in the PPC.
The result follows.
\end{proof}
\end{proposition}

Hence, with fully sophisticated bidders, the PPA second-price auction performs just as well as its PPI and PPC counterparts. This equivalence result may seem at odds with one's intuition that the PPC only prices the value-per-click, but it follows from the fact that $p$ is common to all bidders, and that -- under the \emph{full sophistication} assumption -- all parameters and distributions are perfectly anticipated by all bidders, which (as explained) therefore bid higher than their value-per-click. We discuss next two variations of the baseline model, to accommodate the possibility of heterogeneous attention probabilities, as well as possible framing effects associated to different payment schemes.

\paragraph{Bidder-Specific Attention}

We consider a simple extension of the model, in which the probability of
grabbing a consumer attention varies with the identity of the advertiser. We
thus substitute the $p$ parameter above with a profile $\left(  p_{i}\right)
_{i\in I}$. For each realisation $\left(  p_{i},\gamma_{i}
,q_{i},v_{i}\right)  _{i\in I}$, we relabel advertisers if necessary in decreasing order of value-per-attention and per-impression, respectively: that is, so that $\text{VPA}_{1}\geq
\text{VPA}_{2}\geq...\geq \text{VPA}_{n}$, and $\text{VPI}_{\tilde{1}}\geq
\text{VPI}_{\tilde{2}}\geq...\geq \text{VPI}_{\tilde{n}}$.

\begin{proposition}
For each realisation of types, expected revenues in the dominant-strategy
equilibria of the PPI, PPC and PPA auctions are, respectively, $R^{\text{PPA}}
= R^{\text{PPC}} = p_{1} \text{VPA}_{2}$, and $R^{\text{PPI}} = p_{\tilde{2}} \text{VPA}_{\tilde{2}}$.

\begin{proof}
By the arguments above, $b_{i}^{\text{PPI}} = p_{i}\left[  x\left(  \gamma_{i}-q_{i}\right)  +q_{i}\right]
v_{i}$, $b_{i}^{\text{PPC}} = \left(  \gamma_{i}-q_{i}+\frac{q_{i}}{x}\right)  v_{i}=\text{VPA}_{i}/x$,
and $b_{i}^{\text{PPA}} = \left[  x\gamma_{i}+\left(  1-x\right)  q_{i}\right]  v_{i}=\text{VPA}_{i}$ are dominant strategies.
Therefore, for each $\left(  p_{i},\gamma_{i},q_{i},v_{i}\right)  _{i\in I}$, in the PPI the highest bid $b_{i}^{\text{PPI}}$ is placed by the highest VPI bidder (which is $\hat{1}$), who wins the auction and pays $b_{\tilde{2}}^{\text{PPI}}=\text{VPI}_{\tilde{2}}=p_{\tilde{2}} \text{VPA}_{\tilde{2}}$; in the PPC and PPA auctions instead, the highest bids are placed by the highest VPA bidder (which is $1$), who wins the auction and pays:
(i) $b_{2}^{\text{PPC}} = \text{VPA}_{2}/x$ whenever clicked in the PPC
auction, which happens with probability $p_{1}x$; and (ii) $b_{2}^{\text{PPA}} = \text{VPA}_{2}$
whenever noticed, which happens with probability $p_{1}$. Multiplying the
price by the corresponding probability yields the result.
\end{proof}
\end{proposition}

Thus, for any $\left(  p_{i},\gamma_{i},q_{i},v_{i}\right)  _{i\in I}$, the PPA does just as well as the PPC, and they both outperform the PPI whenever $p_{1}\text{VPA}_{2}\geq p_{\tilde{2}}\text{VPA}_{\tilde{2}}$. The revenue ranking at the ex-ante stage thus depends on the joint distribution of the $p_{i}$'s and $\text{VPA}_{i}$'s. For example, if attention probabilities and values-per attention are positively correlated, the ex-ante expected revenues under the PPA are higher than under the PPI.
This point will be later illustrated with numerical simulations.

\paragraph{Framing Effects}

As explained above, a fully sophisticated advertiser in the PPC should raise their bid by an extra $\left(  1-x\right)  \frac{q_{i}}{x}v_{i}$ over their per-click valuation, because they would take also into account the expected value the ad might generate conditional on not being clicked. This, however, is not a straightforward calculation, and advertisers in practice need not bid this way.
The bidding tutorials provided by most prominent platforms, for example, implicitly assume that $q_{i}
=0$.\footnote{See e.g., the Google AdWords tutorial in which Hal Varian teaches how to bid in the GSP auction: \url{https://www.youtube.com/watch?v=jRx7AMb6rZ0} --- recall that, in the single-slot case, the GSP auction coincides with the PPC second-price auction discussed above.} Hence, if advertisers with $q_{i}>0$ followed the tutorials' recommendation, they would fail to take the term $\left(  1-x\right)  \frac{q_{i}}{x}v_{i}$ into account, and hence they would bid based on the VPC alone, not on their full value-per-impression (VPI).

Bidding tutorials are only one of the reasons why advertisers may bid this way. More broadly, by drawing advertisers' attention to different components of their valuation, it may be that different pricing systems produce \emph{framing effects} which may impact the way advertisers bid in practice, for example by only taking into account the component of the value which is made salient by the particular pricing scheme. These considerations are relevant in practice, since the level of understanding of the environment implicit in the model with fully sophisticated bidders, and the associated calculations needed to obtain the optimal bidding strategies, are unlikely to be completely reflected in the abilities and sophistication of real-world bidders.

For these reasons, we also study the performance of the three auction formats if advertisers were affected by the pricing system, in the sense that they only focus on the value made salient by the rules of the auction, which is the one they pay for: the VPI in the PPI, the VPC in the PPC, and the VPA in the PPA.  For ease of reference, we reproduce here the formulae for these different values:

\begin{align*}
& \text{VPI}_{i} := p_{i} \left[x \gamma_{i} + \left(1 - x\right) q_{i}\right] v_{i} \\
& \text{VPC}_{i} :=\gamma_{i} v_{i} \\
& \text{VPA}_{i} := \left[x \gamma_{i} + \left(1 - x\right) q_{i}\right] v_{i} \\
\end{align*}

As above, for each realisation $\left(  p_{i},\gamma_{i}
,q_{i},v_{i}\right)  _{i\in I}$, we relabel advertisers if necessary in decreasing order of value per-attention, per-click and per-impression, respectively: that is, so that $\text{VPA}_{1}\geq
\text{VPA}_{2}\geq...\geq \text{VPA}_{n}$, $\text{VPC}_{\bar{1}}\geq
\text{VPC}_{\bar{2}}\geq...\geq \text{VPC}_{\bar{n}}$, and $\text{VPI}_{\tilde{1}}\geq
\text{VPI}_{\tilde{2}}\geq...\geq \text{VPI}_{\tilde{n}}$.
\begin{proposition}
If bidders only focus on the value which is made salient by the auction rules,
for each realisation of types, the revenues in the dominant-strategy
equilibria of the PPA, PPC and PPI auctions are, respectively, $\hat{R}^{\text{PPA}} = p_{1}\left[
x\gamma_{2}+\left(  1-x\right)  q_{2}\right]  v_{2}$, $\hat{R}^{\text{PPC}} = p_{\bar{1}}x\cdot\gamma_{\bar{2}}v_{\bar{2}}$ and $\hat{R}
^{\text{PPI}} = p_{\tilde{2}}\left[  x\gamma_{\tilde{2}}+\left(  1-x\right)  q_{\tilde{2}}\right]  v_{\tilde{2}}$. Hence: (i) $\hat{R}^{\text{PPA}} > \hat{R}^{\text{PPI}}$ if
and only if $p_{1}/p_{\tilde{2}}\geq \text{VPA}_{\tilde{2}}/\text{VPA}_{2}$; and (ii) if $p_{1}\geq p_{\bar{1}}$ and $\text{VPA}_{\bar{2}}\neq \text{VPA}_{1}$, then $\hat{R}
^{\text{PPA}} > \hat{R}^{\text{PPC}}$ whenever $q_{\bar{2}}>0$.

\begin{proof}
If bidders only focus on the value which is made salient by the auction rules,
the perceived payoff in case of win (i.e., if $b_{i}>b_{-i}^{\max}$) are equal
to $\hat{u}_{i}^{\text{PPI}}\left(\text{win}\right) = \text{VPI}_{i}-b_{-i}^{\max}$ and $\hat
{u}_{i}^{\text{PPC}}\left(\text{win}\right) = p x\left(\text{VPC}_{i}-b_{-i}^{\max}\right)$,
and $\hat{u}_{i}^{\text{PPI}}\left(\text{win}\right) = \text{VPI}_{i}-b_{-i}^{\max}$, where
$\text{VPC}_{i} = \gamma_{i}v_{i}$, $\text{VPI}_{i} = p_{i}\text{VPA}_{i}$ and $\text{VPA}_{i} = \left[  x\gamma
_{i}+\left(  1-x\right)  q_{i}\right]  v_{i}$. By the usual argument, it is
easy to show that dominant-strategies are, respectively, $\hat{b}_{i}
^{\text{PPI}} = \text{VPI}_{i}$, $\hat{b}_{i}^{\text{PPC}} = \text{VPC}_{i}$ and $\hat{b}_{i}^{\text{PPA}} = \text{VPA}_{i}$.
Hence, for each type realisation, the winner in the PPA, PPC, and PPI is the bidder with the highest VPA, VPC, and VPI, respectively (resp., bidders $i$, $\bar{i}$ and $\tilde{i}$). In the PPA the bidder
pays $\left[  x\gamma_{2}+\left(  1-x\right)  q_{2}\right]  v_{2}$ if noticed; in the PPC he pays $\gamma_{\bar{2}}v_{\bar{2}}$ for each click; in the PPI he
pays $\text{VPI}_{\tilde{2}}=p_{\tilde{2}}\left[  x\gamma_{\tilde{2}}+\left(  1-x\right)  q_{\tilde{2}}\right]
v_{2}$.
The revenues in the statement are obtained multiplying these payments by the
corresponding probabilities ($p_{1}$
in the PPA, $p_{\bar{1}}x$ in the PPC, and $1$ in the PPI). The revenue ranking betwen PPA and PPI follows immediately from these results, noting that revenues can be rewritten as $\hat{R}^{\text{PPA}} = p_{1}\text{VPA}_{2}$ and $\hat{R}^{\text{PPI}} = p_{\tilde{2}}\text{VPA}_{\tilde{2}}$. For the revenue ranking between PPA and PPC, first note that, by definition of the relabellings, $\text{VPA}_{\bar{2}}\neq \text{VPA}_{1}$ implies that $\text{VPA}_{\bar{2}}\leq \text{VPA}_{2}$, and $\text{VPA}_{\bar{2}}>\text{VPC}_{\bar{2}}$ if $q_{\bar{2}}>0$. Hence, if $p_{1}\geq p_{\bar{1}}$, it follows that $\hat{R}^{\text{PPC}}=p_{\bar{1}}x\text{VPC}_{\bar{2}}<p_{\bar{1}}\text{VPA}_{\bar{2}}\leq
p_{\bar{1}}\text{VPA}_{2}\leq p_{1}\text{VPA}_{2} = \hat{R}^{\text{PPA}}$ .
\end{proof}
\end{proposition}

Hence, as long as some advertisers are subject to framing effects in the sense that they solely focus on the value made salient by the rules of the auction, the second-price PPA auction does better than both its PPI and PPC counterparts under weak conditions which are expected to hold in relevant economic settings, as will be illustrated in the numerical simulations of \autoref{sssec:numerical_simulations}.

\section{User Study}
\label{sec:user_study}

Online advertising involves (1)~a publisher who integrates ads into its online content,
such as the native ads on a SERP, promoted tweets on Twitter, or sponsored content in a news stream;
and (2)~an advertiser, who provides the ads to be displayed.
These ads can be served under different formats (e.g., text, image, video, or rich media), each with its unique look and feel. Some formats appear to be more effective than traditional online ads in terms of user attention and purchase intent~\cite{sharethrough2013}, and others may cause ad blindness to a greater or a lesser extent~\cite{Owens:2011:TAB:2007456.2007460}. Therefore, to understand how web search users engage with ads that appear under different formats and positions in SERPs, we conducted a user study through the \textsc{Figure Eight}\footnote{\url{https://www.figure-eight.com}} crowdsourcing platform. Following a similar experimental setup to that introduced by~\citet{Arapakis:2016:PUE:2911451.2911505}, we collected feedback from participants who performed brief transactional search tasks using Google Search. With this study, we aimed to predict when do users notice the ads that appear on SERPs under the aforementioned conditions.

Crowdsourcing studies offer several advantages over in-situ methods of experimentation~\cite{Mason2012},
such as access at a larger and more diverse pool of participants with stable availability,
collection of real usage data at a relatively large scale,
and a low-cost alternative to the more expensive laboratory-based experiments.
On the downside, experimenters have to account for potential threats to ecological validity,
distractions in the physical environment of the participant, and privacy issues, to name a few.
Still, crowdsourcing allows for exploring a wider range of parameters in a more controlled manner
as compared to in-the-wild large-scale studies.
To mitigate and discount low-quality responses, several preventive measures were put into practice,
such as introducing test (gold-standard) questions to our tasks,
selecting experienced contributors (Level 3) with high accuracy rates,
and monitoring their task completion time,
thus ensuring the internal validity of our experiment.

\begin{figure*}[!tpb]
  \def\w{0.32\linewidth}
  \subfloat[Native ad\label{fig:native-top-left}]{
    \includegraphics[trim=0 0 8cm 0, clip=true, width=\w]{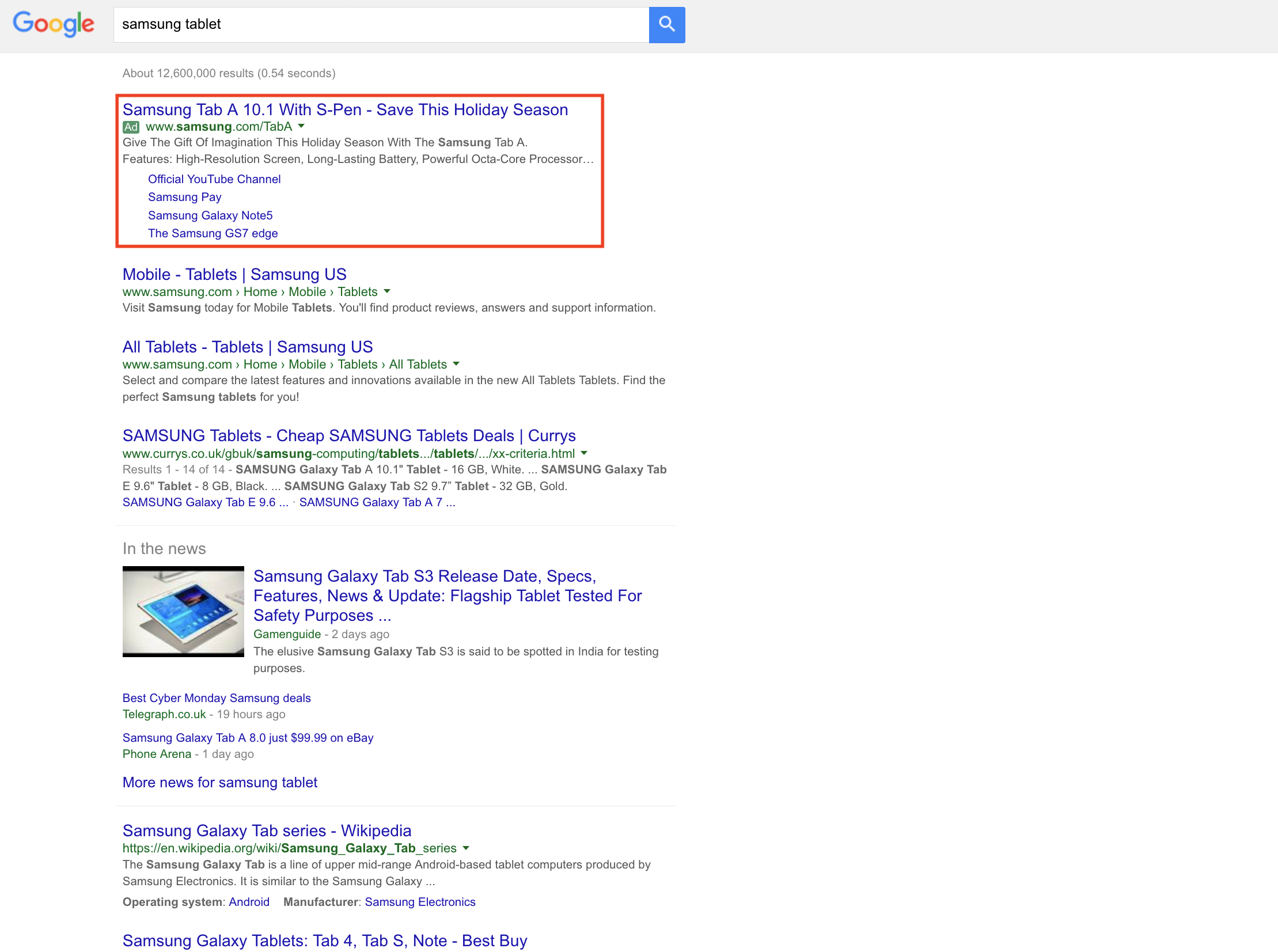}
  }
  \hfill
  \subfloat[Bundled ad (left)\label{fig:dd-top-left}]{
    \includegraphics[trim=0 0 8cm 0, clip=true, width=\w]{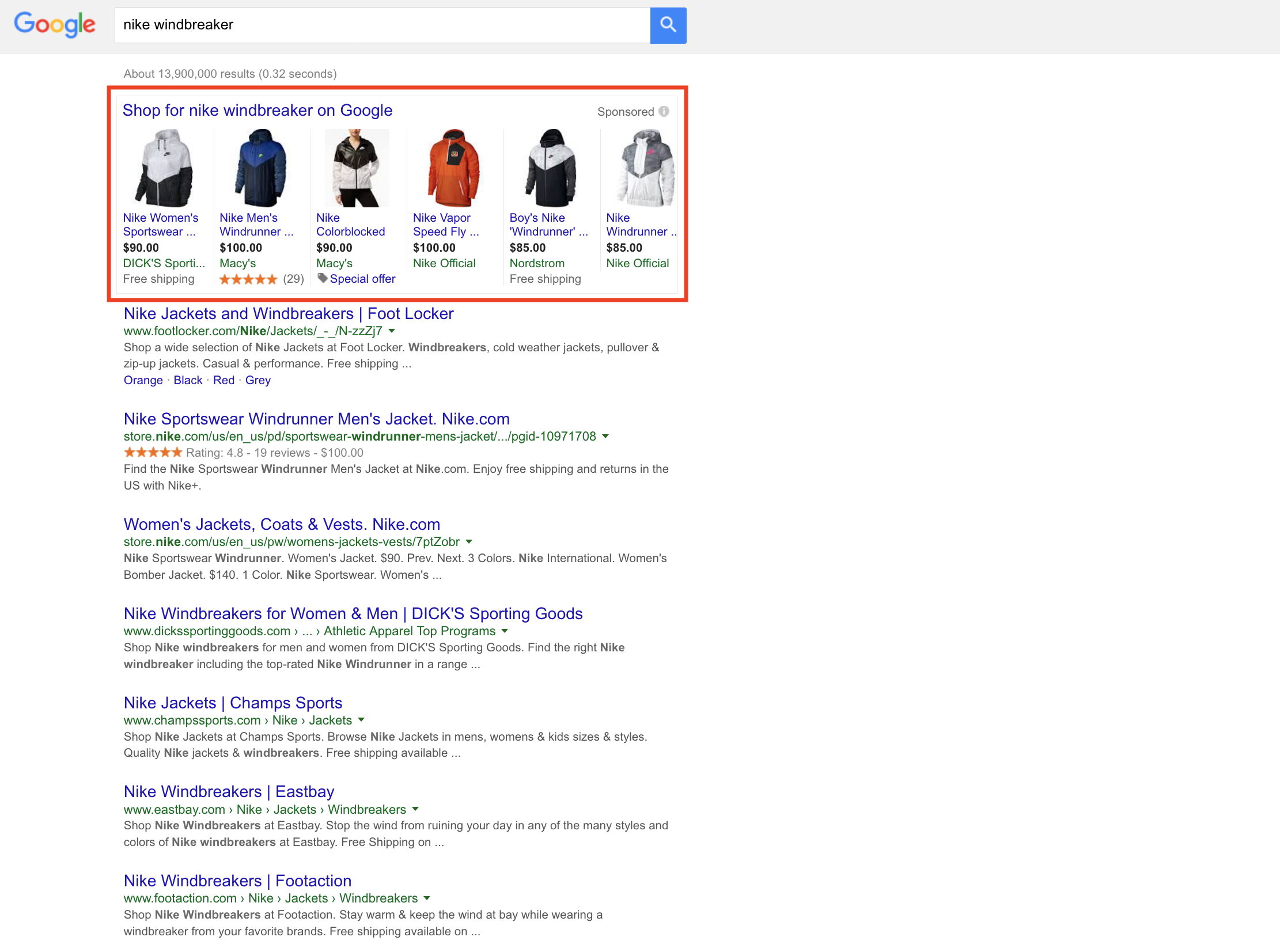}
  }
  \hfill
  \subfloat[Bundled ad (right)\label{fig:dd-top-right}]{
    \includegraphics[trim=0 0 8cm 0, clip=true, width=\w]{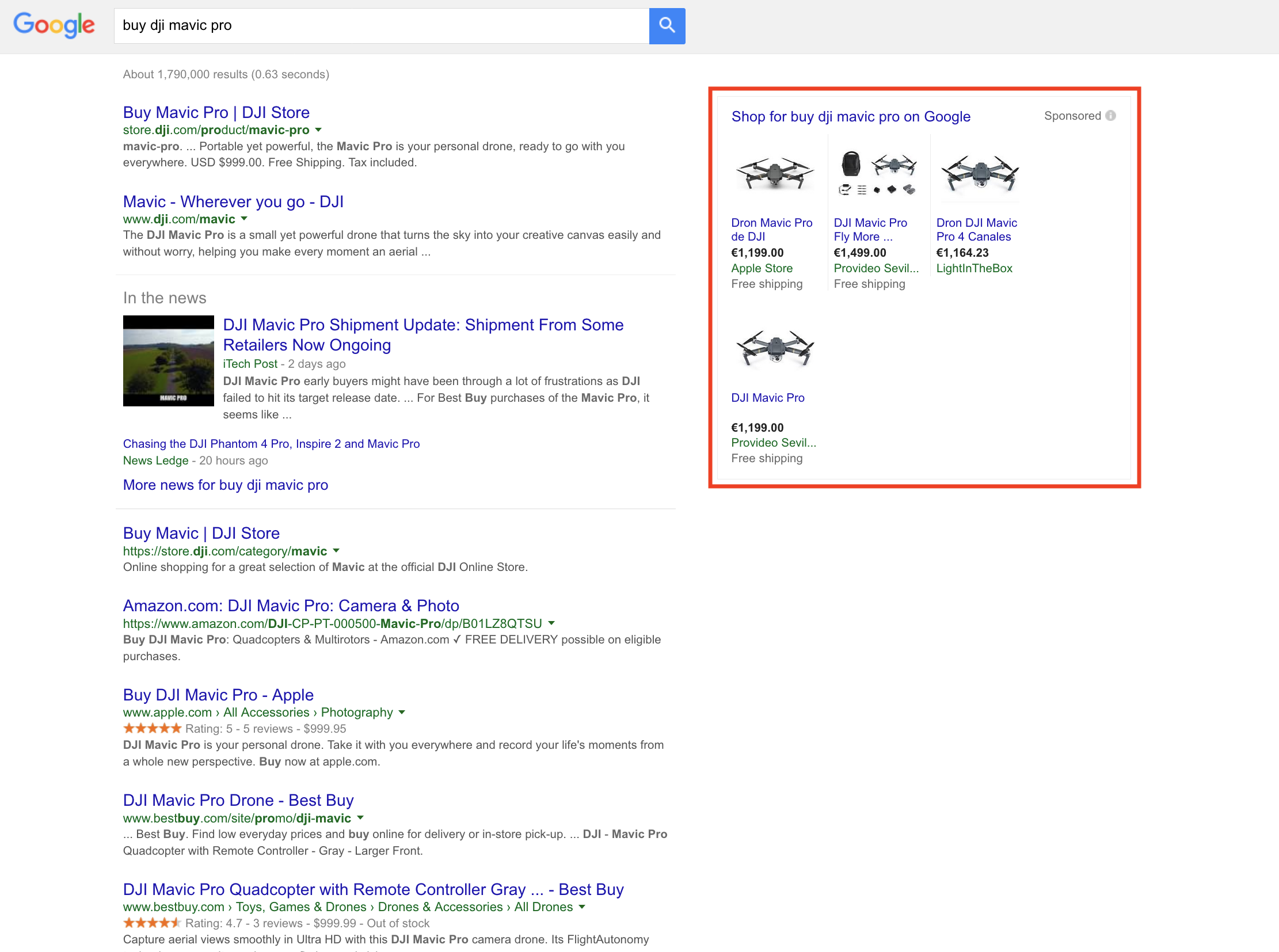}
  }
  \caption{
    Examples of ad formats and their positions on SERPs.
    In our experiments, only one ad format was shown at a time.
  }
  \label{fig:display_ads}
\end{figure*}

\subsection{Experiment Design}
\label{ssec:design}

The experiment had a between-subjects design with two independent variables:
(1)~ad format, with two levels: ``native'' (organic ads) or ``bundled'' (direct display ads),
and (2)~bundled ad position, with two levels: ``left'' and ``right'' position.
Native ads are only shown in the left part of Google SERPs (see below).
The dependent variable was ad attention (see \autoref{sec:user_study}).

Our experiment consisted of a brief transactional search task
where participants were presented with a predefined search query and the corresponding SERP,
and were asked to click on any element of the page that answered it best.
The search queries (\autoref{ssec:search_query_sample}) were all picked from a pool of queries that triggered both native (\autoref{fig:native-top-left}) and bundled ads (Figures~\ref{fig:dd-top-left} and \ref{fig:dd-top-right}) on Google SERPs.
The search queries were randomly assigned to the participants.

All SERPs, which were in English, were scraped for later instrumentation.
As hinted previously, all SERPs had both native and bundled ads.
Native ads appear both at the top-left and bottom-left position of the SERP,
whereas bundled ads could appear either at the top-right or top-left position
(but not both at the same time on the same SERP).
Therefore, we ensured that only one ad was visible per condition and participant at a time,
since we are focusing on the single-slot auction case.
This was possible by instrumenting each downloaded SERP with custom JavaScript code
that removed all ads excepting the one that would be tested in each of the experimental conditions.
In any case, native bottom-most ads were not shown,
since (i)~users have to scroll all way down to the bottom of the SERP to reveal them
and (ii)~these ads have the same look and feel than the native ads shown on the top-most position.

Participants accessed the instrumented SERPs through a dedicated server,
which did not alter the look and feel of the original SERPs.
This allowed us to capture fine-grained user interactions
while ensuring that the content of the SERPs remained consistent
and that each experimental condition was properly administered.
Each participant was allowed to perform the search task only once,
since inquiring at post-task about the presence of an ad would make them aware of it
and could introduce carry over effects, thus altering their browsing behaviour in the subsequent search tasks.
In sum, each participant was exposed only to a single condition;
i.e. a unique combination of query, ad format and ad position.

\subsection{Search Query Sample}
\label{ssec:search_query_sample}

Our search query set was constructed as follows. Starting from \textsc{Google Trends},\footnote{\url{https://trends.google.com/trends/}} we selected a subset of the Top Categories and Shopping Categories (\autoref{tbl:search_query_categories}) that were suitable candidates for the transactional character of our search tasks. From this subset of categories, we extracted the top search queries issued in the US during the last 12 months. Next, from the resulting collection of 375 search queries, we retained 150 for which the SERPs were showing at least one bundled ad (50 search queries for each combination of bundled ad format and position).
Such examples include the search queries \emph{samsung tablet}, \emph{casio watches}, or \emph{adidas ultra boost}. Using this final selection of search queries, we produced the static version of the corresponding Google SERPs and injected the JavaScript code (\autoref{ssec:mouse_cursor_tracking}) that allowed us to control the ads format and capture all client-side user interactions. The final collection of 150 search queries per ad condition was repeated as many times as needed to produce the desired number of search sessions for the final dataset.

\begin{table}[ht!]
\caption{Selected search query categories (Google Trends).}
\label{tbl:search_query_categories}
\centering
{\small
  \begin{tabular}{@{}lll@{}}
  \toprule
  \textbf{Top Categories} & \textbf{Shopping Categories} \\
  \midrule
  Autos \& Vehicles  &   Apparel    \\
  Computers \& Electronics  &   Event Ticket Sales    \\
  Food \& Drink  &   Gifts \& Special Event    \\
  Games  &   Luxury Goods    \\
  Real Estate  &   Photo \& Video Services    \\
  Travel  &   Sporting Goods    \\
    &   Tobacco Products    \\
    &   Toys    \\
    &   Wholesalers \& Liquidation    \\
  \bottomrule
  \end{tabular}
}

\end{table}

\subsection{Mouse Cursor Tracking}
\label{ssec:mouse_cursor_tracking}

As previously stated, all SERPs were downloaded and instrumented with custom JavaScript code.
This way,
we could automatically insert mouse tracking code and log cursor movements, hovers, and associated metadata. For this, we used \textsc{EvTrack},\footnote{\url{https://github.com/luileito/evtrack}} an open source JavaScript event tracking library derived from the smt2$\epsilon$ system~\cite{Leiva:2013:WBB:2540635.2529996}. \textsc{EvTrack} makes it possible to specify \emph{which} browser events should be captured and \emph{how}, i.e., via event listeners (the event is captured as soon as it is fired) or via event polling (the event is captured at fixed-time intervals). Concretely, we captured all regular browser events (e.g., \texttt{\small{load}}, \texttt{\small{click}}, \texttt{\small{scroll}}) via event listeners and only \texttt{\small{mousemove}} via event polling (every 150\,ms), since this event may introduce unnecessary overhead both while recording on the client side and while transmitting the data to the server~\cite{LEIVA2015114}. Whenever an event was recorded, we logged the following information:
mouse cursor position (x and y coordinates), timestamp, event name, xpath of the DOM element that relates to the event, DOM element attributes, and the Euclidean distance to five control points (four corners and middle point) of the ad.
This distance was required by one of the baseline models we tested (\autoref{ssec:baseline}).

\subsection{Self-Reported Measures}
\label{ssec:self_reported_measures}

In addition to the aforementioned mouse cursor data, we collected ground-truth labels on the noticeability of the ads through an online questionnaire. Similar to what other works have done before~\cite{Feild:2010:PSF:1835449.1835458,Liu:2015:DUD:2766462.2767721,Lagun:2014:DCM:2556195.2556265,Arapakis:2016:PUE:2911451.2911505}, the questionnaire was administered at post-task and asked the following question: \emph{While performing the search task, to what extent did you pay attention to the advertisement?}
We used a 5-point Likert-type scale to collect the labels:
1 (``Not at all''), 2 (``Not much''), 3 (``I can't decide''), 4 (``Somewhat''), and 5 (``Very much'').

These scores would be later collapsed to binary labels (true/false),
but we felt it was necessary to use a 5-point Likert-type scale for several reasons.
First, using 2-point scales often results in highly skewed data~\cite{Johnson82}.
Second, it is important to leave room for neutral responses,
because some users may not want to say one way or another, otherwise this can produce response biases.
But 3-point scales can lead more users to stay neutral,
because the remaining options can be seen as ``too extreme''.
Therefore, we opted for a 5-point scale,
which leaves more room for ``soft responses'' and in addition is easy to understand.
With this scoring scheme, therefore, we are confident that
the eventual binary labels actually reflect positive and negative user votes.
All tasks that received a neutral score were not considered for analysis.

\subsection{Participants}
\label{ssec:participants}

We recruited $3,206$ participants through \textsc{Figure Eight},
of which we retained data from $2,289$ (``female'' = $40.9\%$, ``male'' = $58.6\%$, ``Prefer not to say'' = $0.5\%$)
after excluding those cases which had incomplete mouse cursor logs.
Participants aged from 18 to 66
(``18--23'' = $13.6\%$, ``24--29'' = $22.5\%$, ``30--35'' = $21.2\%$, ``36--41'' = $15.6\%$, ``42--47'' = $9.3\%$, ``+48'' = $17.8\%$), were of mixed nationality (e.g., American, Belgian, British, German) and had diverse educational background: 21.6\% had a high school diploma, 16.9\% had a college diploma, 27.1\% had a BSc degree, 17.5\% were graduates, 14.3\% had an MSc, 1.1\% had a PhD, and 1.5\% preferred not to say. The majority were full-time (45.1\%) or part-time (15.38\%) employees while the remaining were either full-time students (11.6\%), pursuing further studies while working (13.9\%), performing home duties (6.5\%) or other (7.5\%). Finally, all participants were proficient in English and were experienced (Level 3) contributors.

\subsection{Procedure}
\label{ssec:procedure}

Initially, the participants were instructed to read carefully the terms and conditions of the study which, among other things, informed them that they should perform the task from a desktop or laptop computer using a computer mouse (and refrain from using a touchpad, tablet, or mobile device) and that they should deactivate any ad-blocker before proceeding with the search task. Our JavaScript code detected any installed ad-blockers and thus prevented the user from taking part in the study.

Participants were also asked to act naturally and choose anything that would best answer the search query, since all ``clickable'' elements (e.g., result links, images, etc.) on the SERP were considered valid answers. The instructions were followed by a brief search task description like ``\emph{Imagine that you want to buy <noun> (for you or someone else as a gift) and you have submitted the search query `<noun>' to Google Search. Please browse the search results page and click on the element that you would normally select under this scenario.}''

The search task had to be completed in a single session and each search query was performed on average by five different participants. The SERPs were randomly assigned to the participants and each participant could take the study only once (see~\autoref{ssec:design}). The participants were allowed as much time as they needed to examine the SERP and proceed with the search task, which would conclude whenever they selected for any of the ``clickable'' elements on the SERP. Upon concluding the search task, participants were asked to complete the post-task questionnaire (which inquired about the presence of the ad and other ground-truth information) and a brief demographics questionnaire. The payment for the participation was \$0.20. Participants could also opt out at any moment, in which case they were not compensated.

\section{Predicting Ad Attention}
\label{sec:predicting_attention}

In this section we present our diagnostic technology
for predicting user attention to ads on SERPs,
using as ground-truth the labels collected in our user study.
To this end, we implement several baseline models:
the Random Forest classifier proposed in~\cite{Arapakis:2016:PUE:2911451.2911505},
a ZeroR classifier that always predicts the majority class,
and a feed-forward neural network using three classic IR features, see \autoref{ssec:baseline}.
We also implement a recurrent neural network exclusively on the raw sequences of 2D mouse cursor coordinates that we collected (\autoref{ssec:rnn_model}). Then, we compare and contrast the accuracy of these models' predictions for the three different ad conditions from our user study: (1) native ad, (2) left-bundled ad and (3) right-bundled ad, and for different demographic attributes (gender, age). Our findings show that our recurrent neural network model achieves better performance over the baseline models in most cases, while avoiding the additional cost of feature engineering and the use of additional page-level information.

\subsection{Data Set}
\label{ssec:data}

After excluding those logs with incomplete mouse cursor data (less than five mouse coordinates, which corresponds roughly to one second of user interaction data), we concluded on a set of $45,082$ cursor positions from $2,289$ search sessions. Of these search sessions, $763$ correspond to the native ad, $793$ correspond to the left-bundled ad, and $733$ correspond to the right-bundled ad. We then converted our ground-truth labels to a binary scale, using the following mapping: (1) ``Not at all'' and (2) ``Not much'' were assigned to the negative class, and (4) ``Somewhat'' and (5) ``Very much'' were assigned to the positive class. We note that the class distribution was fairly balanced (66\% of positive cases) across the experimental conditions. Next, our data set was divided per ad condition, and for each condition we created a 10-fold cross-validation split using stratified sampling to produce balanced splits that preserve the original class distribution. In each fold, 70\% of the data was used for training and 30\% was used for validation.

\subsection{Baseline Models}
\label{ssec:baseline}

We trained a random forest (RF) classifier to predict ad attention that implemented all the features proposed by~\citet{Arapakis:2016:PUE:2911451.2911505}. More specifically, we engineered the base features (e.g., viewport position, cursor distance from the ad, cursor speed, cursor acceleration) and the high-level meta-features (e.g., cursor traversed distance, hovers over the ad, entropy indices, spectral features) derived from the mouse cursor data. \autoref{tbl:features} summarises these features under different categories and also lists the aggregate functions applied to them. We then removed the highly correlated ($r \ge .80,~p < .05$) and linearly dependent features from our feature set. In addition, we normalised the values for all features in the range $[0,1]$ so that feature values that fall in greater numeric ranges would not dominate those in smaller numeric ranges. As a last step, we determined via grid search the optimal hyper-parameter values (number of trees, number of features, $\epsilon$-threshold) for the baseline model and evaluated its performance against the test set.

\begin{table*}[t!]
\caption{Features used by the baseline RF model for predicting ad attention.}
\label{tbl:features}
\centering
{\scriptsize
  \begin{tabular}{@{}lll@{}}
  \toprule
  \textbf{Base features} & \textbf{Meta-features} & \textbf{Aggregate functions}$^*$ \\
  \midrule
  Viewport (width, height)        &  \# Moves (towards, away) Ad                    &  $x_{\min}$, $x_{\max}$  \\
  Cursor positions and timestamps  &  \# Moves (towards, away) Ad within dist. $d$  & $\Sigma$, $\mu$, $\tilde{x}$  \\
  Unique cursor positions          &  \# Clicks (inside, outside) Ad                &  $\sigma^2$, $\sigma_{\chi}$, SST  \\
  Normalised viewport positions    &  \hspace{3.5mm}Time to first click on Ad        &  $\sum$ intra-distances of cursor positions w.r.t. Ad  \\
  Unique normalised viewport pos. &  \# Preceding clicks to Ad                      &  Shannon entropy  \\
  Subsequent points' distance      &  \# Hovers over Ad                              &  Permutation entropy $(w \in \{2, \ldots, 5\}$)  \\
  Subsequent points' duration     &  \# Hovers over other elements                  &  Weighted Permutation entropy $(w \in \{2, \ldots, 5\}$)  \\
  Cursor distance from Ad          &  \# Hovers over Ad vs. other elements          &  Approximate entropy $(w \in \{2, \ldots, 5\}$)  \\
  Cursor speed                    &  \# Preceding hovers over other elements        &  FFT$\colon i_{th}$ most powerful frequency $(i \in \{1, \ldots, 5\}$)\\
  Cursor normalised speed          &  \hspace{3.5mm}Time to first hover (Ad, other elements) &  Multivariate KL div. (symmetric, non-symmetric)  \\
  Cursor acceleration              &  \hspace{3.5mm}Time hovering (Ad, other elements)  &  Earth mover's distance \\
  Cursor normalised acceleration  &  \hspace{3.5mm}Distance traversed overall          &  Hausdorff distance  \\
  Cursor position status wrt. Ad  &  \hspace{3.5mm}Distance traversed (inside, outside) Ad  &  \\
  Vector angles                    &  \hspace{3.5mm}Distance from Ad (corners, center)      &  \\
                                  &  \# Cursor positions within distance $d$ from Ad        &  \\
  \bottomrule
  \\[-0.75em]
  \multicolumn{3}{l}{* These functions are computed for most base and meta-features.}
  \end{tabular}
  }
\end{table*}

We also tested a ZeroR classifier, also known as 0-R (zero rule),
which simply predicts the majority class.
It will always output the same target value and does not use any input features, hence its name.
Despite its simplicity and lack of discriminative power,
this classifier is very useful for determining the baseline performance,
as a benchmark for other classification methods like the ones we used in these experiments.
If any other classifier is correct less frequently than ZeroR,
it is obviously of no value for the task at hand.

\subsection{Feed-forward Neural Network Model}
\label{ssec:ann_model}

The RF model introduced previously is a machine learning ensemble
using a sum of piecewise functions for classification,
therefore it may have limited accuracy.
In contrast, neural networks can easily model any dependencies within the data.
Therefore, we trained a feed-forward neural network (FFNN) as a third baseline model.
The FFNN uses three classic features from the literature
that have been suggested to correlate well with user engagement~\cite{Arapakis:2014:UEO:3151365.3151368, Barbieri:2016:IPU:2872427.2883092, Lagun:2015:ISA:2766462.2767745}:
dwell time, number of clicks over the ad, and number of hovers over the ad.

The FFNN input layer takes a vector with these three features
and feeds it to a fully-connected hidden layer
with 6 neurons (two neurons per feature) and ReLU activation,
followed by a dropout layer with drop rate $q=0.5$ for regularisation,
and finally a fully-connected layer as output with 1 neuron and sigmoid activation.
The FFNN outputs a probability prediction $p$ of the user's attention to an ad,
where $p>.5$ indicates that the user has noticed the ad.

We trained the FFNN with a batch size of 64 sequences and for 50 epochs,
using the same 10-fold cross-validation splits as the RF model.
We used the popular Adam optimizer (stochastic gradient descent with momentum)
with learning rate $\eta=0.001$ and decay rates $\beta_1=0.9$ and $\beta_2=0.999$.
The loss function to minimise is binary cross-entropy, since the task is a 2-class classification problem.

\subsection{Recurrent Neural Network Model}
\label{ssec:rnn_model}

Feature engineering requires domain expertise to come up with the optimal set of discriminative features.
The previous baseline models use ad-hoc features
that exploit the SERP structure and thus are potentially less generalizable.
Given that we are interested in a scalable diagnostic technology of user attention,
regardless the underlying page contents or structure,
we propose a more versatile model to predict user attention to ads.
The model is a particular type of recurrent neural networks (RNNs),
since mouse movements are of sequential nature
and RNNs are very good at modelling data sequences and time series;
were each multivariate data point can be assumed to be dependent on the previous ones.

\begin{figure}[!ht]
    \centering
    \includegraphics[width=0.8\linewidth]{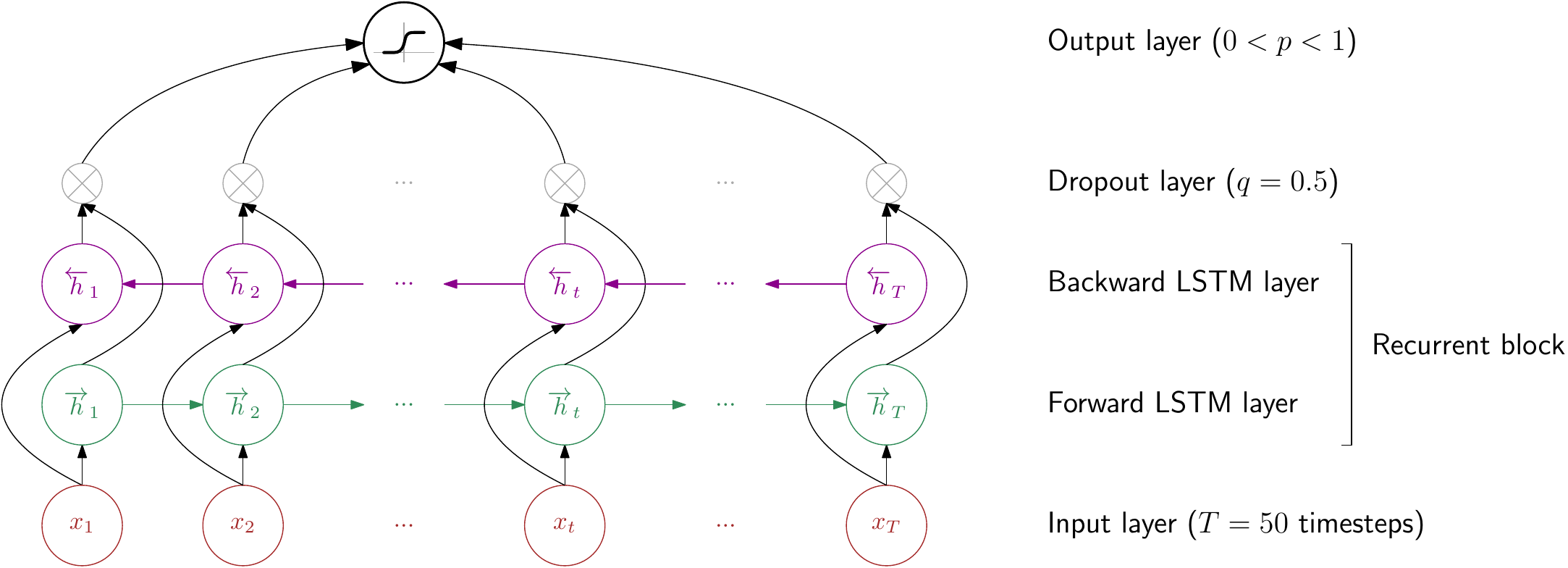}
    \caption{Diagram of our bidirectional LSTM architecture.}
    \label{fig:blstm}
\end{figure}

Concretely, the model architecture is a bidirectional long short-term memory (LSTM) network; see \autoref{fig:blstm}.
An LSTM network is essentially an RNN that can remember long-term dependencies.
We used the bidirectional variant (BLSTM) since a major issue with all RNNs is that they can only learn representations from \emph{previous} time steps. However, sometimes we have to learn representations from \emph{future} time steps to better understand the context and thus eliminate potential ambiguities.

The BLSTM takes as input a sequence of raw mouse cursor positions,
which can be seen as a multivariate time series of two-dimensional data points.
The input layer has 50 neurons (one neuron per timestep).
The hidden layer is a recurrent block with a forward + backward LSTM,
with hyperbolic tangent as activation function
and sigmoid activation in the recurrent step.
Similar to the previously discussed FFNN model,
we added a dropout layer with drop rate $q=0.5$ for regularisation,
followed by a fully-connected layer of 1 output unit using sigmoid activation.
The BLSTM outputs a probability prediction $p$ of the user's attention to an ad,
where $p>.5$ indicates that the user has noticed the ad.
In sum, the only architectural difference between our FFNN and BLSTM models
is the input layer and the first hidden layer.
The model architecture is illustrated in \autoref{fig:blstm}.

Since our BLSTM takes as input a raw sequence of mouse cursor positions only,
and because each sequence has a different length,
the input sequences are padded to a fixed length of 50 timesteps,
which corresponds roughly to the mean sequence length observed in our data set plus one standard deviation.
Also because each mouse cursor sequence was performed on different web browsers
with different screen sizes and thus different positions of the SERP components,
horizontal coordinates were normalised by each user's viewport width.

We trained this model with a batch size of 64 sequences and for 50 epochs,
using the same 10-fold cross-validation splits as the baseline models.
We used the popular Adam optimizer (stochastic gradient descent with momentum)
with learning rate $\eta=0.001$ and decay rates $\beta_1=0.9$ and $\beta_2=0.999$.

\subsection{Results}
\label{ssec:resutls}

\subsubsection{Classification Accuracy}
\label{sssec:baseline_vs_blstm}

\autoref{tbl:attention} shows the experimental results comparing the baseline models~(\autoref{ssec:baseline})
and our recurrent neural network~(\autoref{ssec:rnn_model}) of user attention.
We report weighted Precision, Recall, and F-measure (F1 score),
according to the target class distributions in each case, averaged across the ten folds.
We also report the Area Under the ROC curve (AUC),
to highlight the discriminative power of each classifier.
We use the Pearson's $\chi^2$ test of proportions as omnibus test
and, if the result of the omnibus test reveals a statistically significant difference,
we use as post-hoc test pairwise comparisons between pairs of proportions with correction for multiple testing,
to see if a statistically significant difference exists between individual models.

\bgroup
\def\arraystretch{1.1}
\begin{table}[!ht]
    \caption{Ad attention prediction results, weighted by class distribution. A bold typeface denotes the best result for the corresponding experimental condition.}
    \label{tbl:attention}
    \centering
    {
    \begin{tabular}{*6l}
      \toprule
      \textbf{Ad condition}
        & \textbf{Model}
        & \textbf{Adj. Precision}
        & \textbf{Adj. Recall}
        & \textbf{Adj. F-measure}
        & \textbf{AUC} \\
      \midrule
        Native          & RF        & 0.584     & 0.570     & 0.568     & 0.601 \\
                        & ZeroR     & 0.465     & 0.682     & 0.553     & 0.500 \\
                        & FFNN      & 0.515     & 0.663     & 0.552     & 0.473 \\
                        & BLSTM     & \bf 0.712 & \bf 0.714 & \bf 0.650 & \bf 0.634 \\
      \midrule
        Bundle, left    & RF        & \bf 0.592 & 0.580     & 0.578     & \bf 0.641 \\
                        & ZeroR     & 0.524     & \bf 0.724 & \bf 0.608 & 0.500 \\
                        & FFNN      & 0.524     & \bf 0.724 & \bf 0.608 & 0.519 \\
                        & BLSTM     & 0.524     & \bf 0.724 & \bf 0.608 & 0.624 \\
      \midrule
        Bundle, right   & RF        & 0.537     & 0.514     & 0.498     & 0.590 \\
                        & ZeroR     & 0.485     & \bf 0.692 & 0.570     & 0.500 \\
                        & FFNN      & 0.485     & \bf 0.692 & 0.570     & 0.501 \\
                        & BLSTM     & \bf 0.560 & 0.687     & \bf 0.576 & \bf 0.630 \\
      \bottomrule
    \end{tabular}
    }
\end{table}
\egroup

Our findings indicate that our BLSTM classifier achieves competitive performance in detecting ad attention,
as compared to the other models, for most metrics and under most ad conditions.
Notice that the BLSTM does not use engineered features (like the RF model)
and does not use page-level information (like the FFNN model),
only the raw sequences of mouse movement coordinates.

For the case of native ads, the omnibus test was statistically significant for all metrics.
In terms of Precision\\ \statreport{\chi^2_{(3, N=712)} = 99.891, p < .0001, \phi = 0.375},
BLSTM performed significantly better than the other models,
and there was no statistically significant difference between ZeroR and FFNN.
All other differences were statistically significant.
The effect size suggests a moderate practical importance.
In terms of Recall \statreport{\chi^2_{(3, N=712)} = 36.280, p < .0001, \phi = 0.226},
all models performed better than RF.
All other differences were not statistically significant.
The effect size suggests a moderate practical importance.
In terms of F-measure \statreport{\chi^2_{(3, N=712)} = 19.168, p < .001, \phi = 0.164},
BLSTM performed significantly better than the other models.
All other differences were not statistically significant.
The effect size suggests a small practical importance.
In terms of AUC \statreport{\chi^2_{(3, N=712)} = 52.026, p < .0001, \phi = 0.270},
both BLSTM and RF performed significantly better than FFNN and ZeroR,
and the difference between BLSTM and RF was not statistically significant.
The difference between FFNN and ZeroR was not statistically significant.
The effect size suggests a moderate practical importance.

For the case of left-bundled ads, the omnibus test was statistically significant for all metrics
excepting F-measure \statreport{\chi^2_{(3, N=748)} = 2.105, p = .5510, \phi = 0.053}.
The effect size suggests a small practical importance.
In terms of Precision \statreport{\chi^2_{(3, N=748)} = 10.446, p = .0151, \phi = 0.118},
the post-hoc tests revealed no statistically significant differences between models.
The effect size suggests a small practical importance.
In terms of Recall \statreport{\chi^2_{(3, N=748)} = 54.193, p < .0001, \phi = 0.269},
all models performed better than RF.
All other differences were not statistically significant.
The effect size suggests a moderate practical importance.
In terms of AUC \statreport{\chi^2_{(3, N=748)} = 47.190, p < .0001, \phi = 0.251},
both BLSTM and RF performed significantly better than FFNN and ZeroR,
and the difference between BLSTM and RF was not statistically significant.
The difference between FFNN and ZeroR was not statistically significant.
The effect size suggests a moderate practical importance.

For the case of right-bundled ads, the omnibus test was statistically significant for all metrics.
In terms of Precision \statreport{\chi^2_{(3, N=692)} = 11.907, p < .01, \phi = 0.131},
BLSTM performed significantly better than FFNN and ZeroR,
and the difference between BLSTM and RF was not statistically significant.
All other differences were not statistically significant.
The effect size suggests a small practical importance.
In terms of Recall \statreport{\chi^2_{(3, N=692)} = 70.640, p < .0001, \phi = 0.320},
BLSTM performed significantly better than the other models.
All other differences were not statistically significant.
The effect size suggests a moderate practical importance.
In terms of F-measure \statreport{\chi^2_{(3, N=692)} = 11.567, p < .01, \phi = 0.129},
the RF performed significantly worse than the other models.
All other differences were not statistically significant.
The effect size suggests a small practical importance.
In terms of AUC \statreport{\chi^2_{(3, N=692)} = 35.842, p < .0001, \phi = 0.228},
both BLSTM and RF performed significantly better than FFNN and ZeroR,
but the difference between BLSTM and RF was not statistically significant.
The difference between FFNN and ZeroR was not statistically significant.
The effect size suggests a moderate practical importance.

The case of left-bundled ads is interesting.
As can be observed in \autoref{tbl:attention}, both neural network models
achieved the same Precision, Recall, and F-Measure
as the ZeroR classifier, suggesting that they were unable to model the data distribution
and thus learned to use the prior probability for classification.
The results suggest therefore that predicting attention to left-bundled ads is a challenging task.
Still, the superiority of the BLSTM was evident with respect to the AUC.
The AUC represents the capability of a classifier to distinguish between classes,
and when AUC is $0.5$, it means the model has no class separation capacity,
as is the case of the ZeroR classifier.
Thus, we conclude that it is possible to detect user attention to online ads with competitive accuracy.
More importantly, it is possible to do so unobtrusively and at large scale.

In what follows, we examine the effect that certain demographic attributes like gender and age may have on the proposed diagnostic technology of ad attention. Such effects are important as they allow for market segmentation, better ads tailoring, and informing the online auction schemes and further improving the auction performance in various ways.
We use the BLSTM model since it is the best performer overall,
as indicated by the results discussed above.

\subsubsection{Gender Analysis}
\label{sssec:gender}

We were interested in observing how the accuracy of our ad attention model may vary per user gender,
i.e. whether users of a specific gender exhibit more (or less) predictable patterns of attention to ad displays.
To this end, we analysed the mouse cursor data separately for male and female users using the same experimental setup as in \autoref{sssec:baseline_vs_blstm}, and compared the BLSTM model's performance across all ad conditions.
Pearson's $\chi^2$ test with Yates' continuity correction
to assess the differences in accuracy performance
and highlight those cases where the model performs significantly better.

\bgroup
\def\arraystretch{1.1}
\begin{table}[!ht]
    \caption{Variation of ad attention prediction performance by gender.
    `N' denotes the sample size of each group
    and  `Ratio` denotes the number of positive:negative instances (users who noticed vs not noticed the ad).
    A bold typeface denotes the best result.}
    \label{tbl:gender}
    \centering
    {\small
    \begin{tabular}{lrrrrrr}
        \toprule
        \textbf{Gender}
        & \textbf{N}
        & \textbf{Ratio}
        & \textbf{Adj. Precision}
        & \textbf{Adj. Recall}
        & \textbf{Adj. F-measure}
        & \textbf{AUC} \\
        \midrule
        Male    & 1256 & 334:922 & 0.490     & \bf 0.700 & 0.576     & \bf 0.596 \\
        Female  &  884 & 289:595 & \bf 0.652 & 0.691     & \bf 0.593 & 0.576 \\
        \bottomrule

    \end{tabular}
    }
\end{table}
\egroup

The BLSTM model achieved significantly better Precision when predicting attention from female users
than male users \statreport{\chi^2_{(1, N=2140)} = 15.179, p = .0001, \phi = 0.084}.
However, no statistically significant differences were observed for any of the other metrics
and effect sizes were small in all cases \statp{\phi < 0.1},
suggesting thus a small practical importance.
Therefore, we cannot conclude that user's gender plays an important role
in predicting user attention to online ads and, subsequently,
gender should not be used to inform the online auction.

\subsubsection{Age Analysis}
\label{sssec:age}

We were also interested in observing how the accuracy of our diagnostic technology may be affected by users at different ages, i.e. whether users of a specific age group exhibit more (or less) predictable patterns of attention to ad displays, compared to other age groups. We divide our users into six age groups: ``18–23'', ``24-29'', ``30-35'', ``36-41'', ``42-47'', and ``48+''. The age groups are split in such a way that each group has enough users while preserving common understanding of young, adults, middle-aged and elder people.

Again, we use the same setup as in the previous experiments
and compared the BLSTM model's performance across all ad conditions.
We use the Pearson's $\chi^2$ test of proportions as omnibus test
and, if the result of the omnibus test reveals a statistically significant difference,
we use as post-hoc test pairwise comparisons between pairs of proportions with correction for multiple testing,
to see if a statistically significant difference exists between individual age groups.

\bgroup
\def\arraystretch{1.1}
\begin{table}[!ht]
    \caption{Variation of ad attention prediction performance by age.
    `N' denotes the sample size of each group
    and  `Ratio` denotes the number of positive:negative instances (users who noticed vs not noticed the ad).
    A bold typeface denotes the best result.}
    \label{tbl:age}
    \centering
    {\small
    \begin{tabular}{lrrrrrr}
        \toprule
        \textbf{Age Group}
        & \textbf{N}
        & \textbf{Ratio}
        & \textbf{Adj. Precision}
        & \textbf{Adj. Recall}
        & \textbf{Adj. F-measure}
        & \textbf{AUC} \\
        \midrule
        18--23  & 289 &  91:198 & \bf 0.787 & 0.689     & 0.574     & 0.469 \\
        24--29  & 471 & 131:340 & 0.602     & 0.654     & 0.530     & \bf 0.616 \\
        30--35  & 459 & 119:340 & 0.679     & \bf 0.739 & \bf 0.641 & 0.586 \\
        36--41  & 343 & 105:238 & 0.607     & 0.660     & 0.541     & 0.578 \\
        42--47  & 206 &  64:142 & 0.659     & 0.709     & 0.637     & 0.580 \\
        +48     & 383 & 118:265 & 0.531     & 0.721     & 0.612     & 0.598 \\
        \bottomrule

    \end{tabular}
    }
\end{table}
\egroup

There was a statistically significant difference between groups for all metrics.
In terms of Precision\\ \statreport{\chi^2_{(5, N=2151)} = 138.774, p < .0001, \phi = 0.254},
differences between the ``+48'' group and all the other age groups
were found to be statistically significant.
The difference between the ``18--23'' group and any of the remaining groups
and were also statistically significant.
The effect size suggests a small practical importance.
In terms of Recall \statreport{\chi^2_{(5, N=2151)} = 163.304, p < .0001, \phi = 0.276}
and F-measure \statreport{\chi^2_{(5, N=2151)} = 154.792, p < .0001, \phi = 0.268},
the post-hoc tests revealed no statistically significant differences between age groups.
The effect size suggests a small practical importance for both metrics.
In terms of AUC \statreport{\chi^2_{(5, N=2151)} = 160.928, p < .0001, \phi = 0.274},
the difference between the ``18--23'' group and all the other groups
was found to be statistically significant.
The effect size suggests a small practical importance.

As can be observed in \autoref{tbl:age},
the more accurate results were observed for younger age groups, up to 30--35 years old.
Then, as user's age increased, the BLSTM decreased significantly in Precision and consistently increased in Recall.
This observation, together with the fact that the AUC also deteriorated with older age groups,
suggests an increase in the number of false positives,
and therefore a degadation in classification performance.

Our findings underline potential age effects on the way a mouse device is used in an online search task.
We also found that the number of mouse movements consistently increased with age in our dataset;
e.g. the average number of coordinates in the ``18--23'' group is $M=16.97$ ($SD=13.10$),
in the ``30--25'' group is $M=17.69$ ($SD=15.17$),
and in the ``+48'' group is $M=24.39$ ($SD=24.96$).
However, we should point out that the number of mouse movements alone provides an incomplete picture of age-related effects.
Overall, ageing is marked by a decline in motor control abilities,
therefore it is expected to affect the users' pointing performance
and, by extension, \emph{how} they move the computer mouse.
For example, \citet{Smith1999} observed that older people incurred in longer mouse movement times,
which we also found in our data,
but also more sub-movements and more pointing errors than the young.
Prior work~\cite{HSU1999461,Bohan1998,Jastrzembski2003InputDF,LINDBERG2006170,Smith1999,Walker1997}
has also linked age with motor control and pointing performance in tasks that involve the use of a computer mouse.
Therefore, we conclude that user's age plays an important role
in predicting user attention to online ads and, subsequently,
age could be used to inform the online auction.
We elaborate more about these observations in \autoref{sec:discussion}
and discuss how they may impact our diagnostic technology.

\section{PPA Evaluation}
\label{sssec:numerical_simulations}

Having shown that user attention to ads on SERPs can be accurately predicted and at a large scale,
we proceed to evaluate the expected performance of the PPA auction scheme.
Since we do not have full control over an ads platform to test our method live,
in this section we illustrate our theoretical findings with a series of numerical simulations,
to show how the revenue rankings between the three auction formats are affected
when we vary both the correlation between values and attention probabilities,
and the fraction of bidders subject to framing effects.
We also exemplify the main insights highlighted by our theoretical results
and show their significance in the context of the distribution of attention probabilities
derived from the BLSTM model discussed in the previous section.

First, we assume that parameters $\gamma_{i}$ and $q_{i}$ in \autoref{ssec:environment}
are i.i.d. draws from a uniform distribution over the unit interval,
and that attention probabilities $p_{i}$ are independently drawn from some distribution $P$.
Recall that $\gamma_{i} \geq 0$ denotes the probability that bidder $i$
realizes a sale conditional on the consumer having clicked on the ad
and $q_{i}$ denotes the probability that $i$ realizes a sale conditional on the ad being noticed but not clicked.

To illustrate the effects of varying the correlation between valuations and attention probabilities, we follow the simple statistical model of \citet{AusubelBaranov2018}. Namely, we assume that with probability $(1-\rho)$ valuations are independent draws from a uniform over $[0,100]$; with probability $\rho$ if valuations instead are perfectly correlated with attention probabilities (specifically, such that $v_{i}=p_{i} \cdot 100$). Recall that $v_{i}$ is not a probability, but it denotes $i$'s value for making a sale.\footnote{We normalise the valuations to the interval $[0,100]$ so that revenues in these simulations are directly expressed as percentages of the highest value of making a sale.}
This is a simple statistical model to illustrate how results are affected when one varies the correlation between attention probabilities and values, captured by the $\rho$ parameter (if $\rho=0$, all variables are independent; if $\rho=1$, $p_{i}$ and $v_{i}$ are perfectly correlated). We also let $\alpha\in\left[  0,1\right]  $ denote the fraction of bidders who are subject to the framing effects discussed above.

We then simulate the expected revenues of the three auction formats, assuming that for each $(\gamma_{i}, q_{i}, p_{i}, v_{i} )$ drawn from such a joint distribution, advertisers follow the optimal strategies identified in our results above. Namely, $b_{i}^{PPA}=\left[
x\left(  \gamma_{i}-q_{i}\right)  +q_{i}\right]  v_{i}$, $b_{i}^{\text{PPI}}=p_{i}\left[
x\left(  \gamma_{i}-q_{i}\right)  +q_{i}\right]  v_{i}$ and $b_{i}
^{\text{PPC}}=\left(  \gamma_{i} - q_{i} + \frac{q_{i}}{x}\right)  v_{i}$ if they are sophisticated, and $\hat{b}_{i}^{PPA}=\text{VPA}_{i}$, $\hat{b}_{i}^{\text{PPI}}=\text{VPI}_{i}$ and $\hat{b}_{i}
^{\text{PPC}}=\text{VPC}_{i}$ if they are subject to framing effects.
We thus compute the expected revenues generated by the optimal strategy profiles in the three auction formats (with and without framing effects), by sampling the parameters $(\gamma_{i}, q_{i}, p_{i}, v_{i} )$ from the same common distribution, held constant across the auction formats. The two simulations only differ in the exact specification of such a distribution, and particularly in that of the distribution $P$ of attention probabilities.

\begin{figure*}[!ht]
  \def\w{0.27\textwidth}
  \subfloat[No framing effects\label{fig:no_framing1}]{
    \includegraphics[trim=40px 0px 40px 8px, clip=true, width=\w]{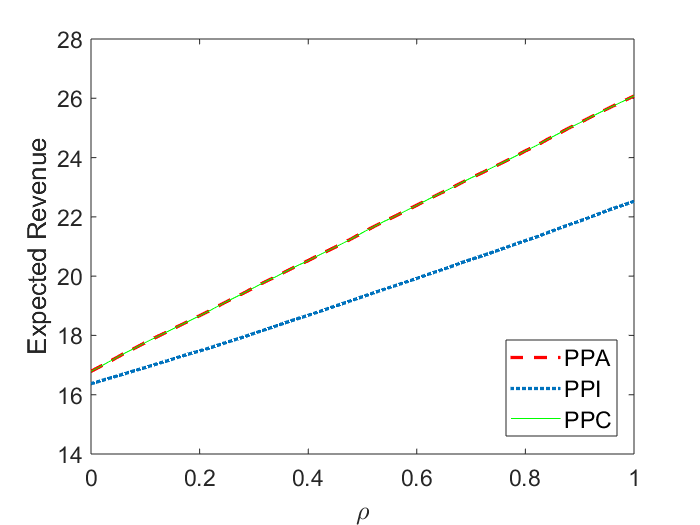}
  }
  \hfill
  \subfloat[Framing effects for half the bidders \label{fig:half_framing1}]{
    \includegraphics[trim=40px 0px 40px 8px, clip=true, width=\w]{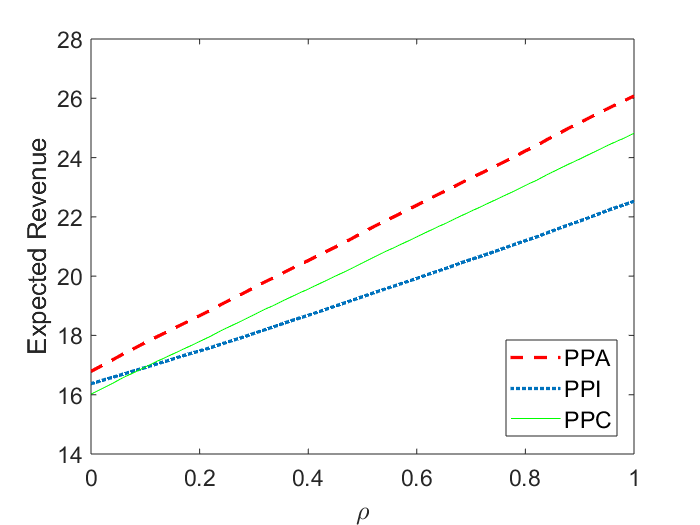}
  }
  \hfill
  \subfloat[Framing effects for all bidders\label{fig:all_framing1}]{
    \includegraphics[trim=40px 0px 40px 8px, clip=true, width=\w]{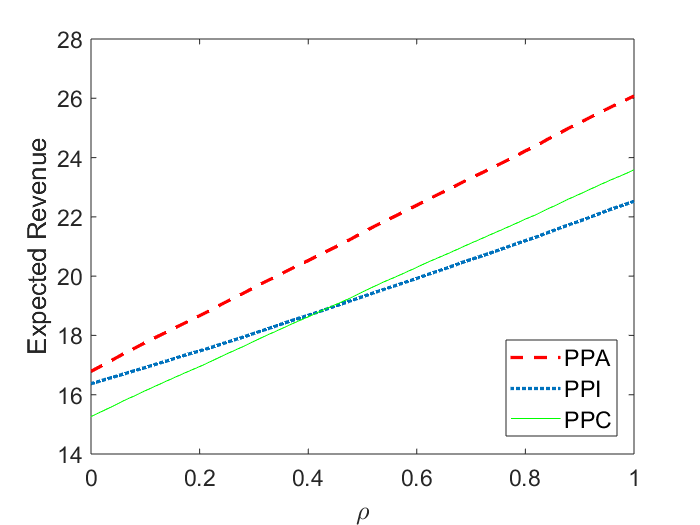}
  }
  \caption{
    Comparisons of expected revenues as a function of the correlation between valuations and attention probabilities,
    with attention probabilities uniformly distributed over the unit interval,
    under varying fractions of bidders subject to framing effects:
    none~\protect\subref{fig:no_framing1},
    half of the bidders~\protect\subref{fig:half_framing1},
    all bidders~\protect\subref{fig:all_framing1}.
  }
  \label{fig:numerical_simulations1}
\end{figure*}

In the first simulation, we set the distribution $P$ of attention probabilities equal to a uniform distribution over $[0,1]$. This textbook example is best suited to illustrate the various possibilities indicated by the theoretical results. Namely, expected revenues under the three auction formats are the same if all bidders are sophisticated ($\alpha=0$) \emph{and} if there is no correlation between valuations and attention probabilities ($\rho=0$); but as soon as $\rho>0$, the difference between the revenues of the PPA and PPI auction becomes positive and increasing in $\rho$. If $\alpha>0$, PPI revenues are strictly higher than those of the PPC for all $\rho \geq 0$, and also higher than those of the PPI for all $\rho >0$, and more so as the fraction $\alpha$ of non-fully sophisticated bidders increases. As for the relative ranking between PPI and PPC revenues, for any $\alpha>0$, there is a $\bar{\rho}(\alpha)$ increasing in $\alpha$ such that revenues are higher in the PPI if $\rho<\bar{\rho}(\alpha)$, and in the PPC if $\rho>\bar{\rho}(\alpha)$. The results are illustrated in \autoref{fig:numerical_simulations1}, respectively, for $\alpha\in\{0, 0.5, 1\}$. For $\alpha=0.5$, for example, PPA revenues are about $1.6\%$ higher than PPC and up to $9\%$ higher than PPI (if $\rho=1$, or about $8\%$ if $\rho=0.5$).

\begin{figure*}[!ht]
  \def\w{0.27\textwidth}
  \subfloat[No framing effects\label{fig:no_framing2}]{
    \includegraphics[trim=40px 0px 40px 8px, clip=true, width=\w]{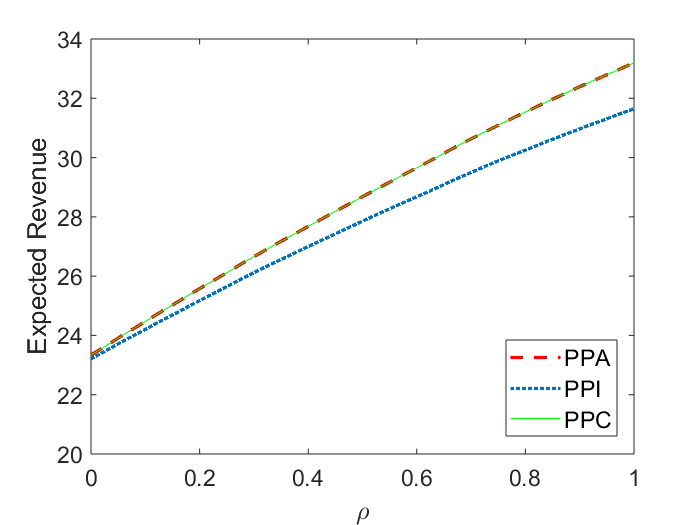}
  }
  \hfill
  \subfloat[Framing effects for half the bidders \label{fig:half_framing2}]{
    \includegraphics[trim=40px 0px 40px 8px, clip=true, width=\w]{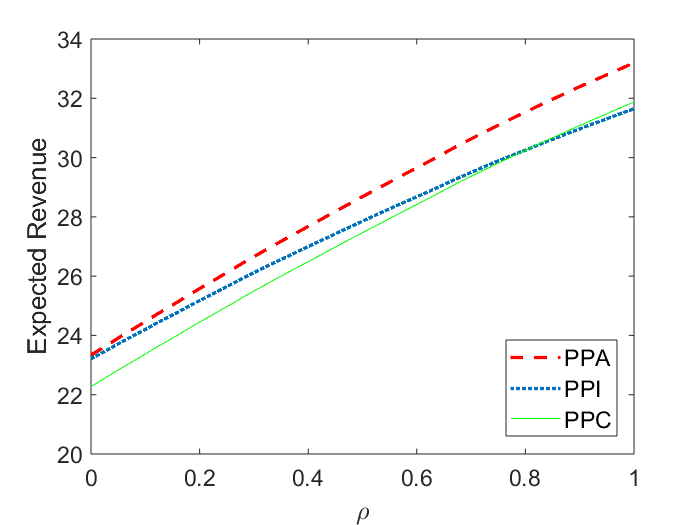}
  }
  \hfill
  \subfloat[Framing effects for all bidders\label{fig:all_framing2}]{
    \includegraphics[trim=40px 0px 40px 8px, clip=true, width=\w]{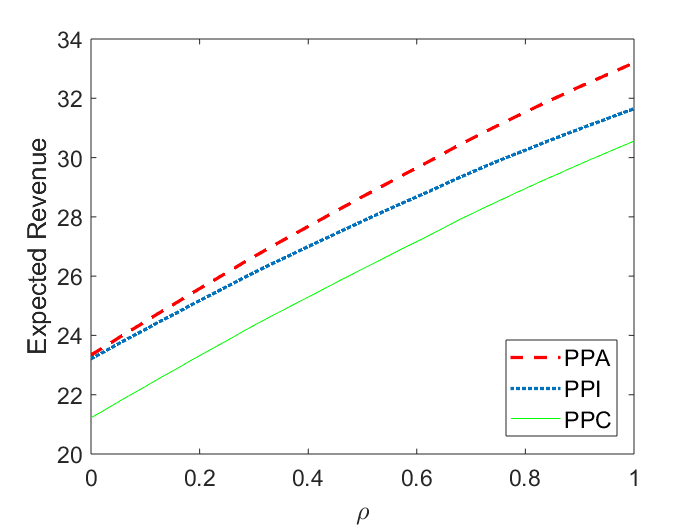}
  }
  \caption{
    Comparisons of expected revenues as a function of the correlation between valuations and attention probabilities,
    with attention probabilities following a $\text{Beta}(15.07;6.65)$ distribution
    (MLE computed from the BLSTM model and the data gathered from our user study),
    under varying fractions of bidders subject to framing effects:
    none~\protect\subref{fig:no_framing2},
    half of the bidders~\protect\subref{fig:half_framing2},
    all bidders~\protect\subref{fig:all_framing2}.
  }
  \label{fig:numerical_simulations2}
\vspace{-0.2cm}
\end{figure*}

In the second simulation, we set the distribution of attention probabilities
equal to a parametric maximum-likelihood estimate (MLE) based on the BLSTM model predictions from our crowdsourced user study.
We note that as long as a diagnostic technology like ours is correct on average -- that is, without systematically under- or over-estimating the attention probabilities -- then the accuracy of the estimates would not affect the optimal bids of risk-neutral advertisers, nor the expected revenues of the auctions. For this reason, the accuracy of the estimated attention probabilities plays no role in these simulations.

More specifically, in the second simulation we fit a Beta distribution to the observed attention probabilities derived from the BLSTM model
and then perform a two-parameter MLE, by pooling data across all subjects and ad types in our experiment.
It is possible to perform alternative simulations using MLE-distributions based on subsamples of the data.
The resulting parameters of the fitted Beta distribution are, respectively, $a=15.07$
and $b=6.65$.
We thus set the distribution of attention probabilities in the numerical simulation equal to $\text{Beta}(15.07;6.65)$
and generate the expected revenues for the three auction formats as explained above.
The results are illustrated in \autoref{fig:numerical_simulations2}, respectively for $\alpha\in\{0, 0.5, 1\}$.

Comparing these results with those from the previous simulations, it is interesting to note that using the MLE of the observed probabilities (derived from the BLSTM model) increases the extent by which the PPA outperforms the PPC.
For example, for $\alpha=0.5$, PPA revenues are about $5\%$ higher than those of the PPC across all values of $\rho$. In contrast, the difference between revenues under the PPA and PPI for low values of $\rho$ is less pronounced than in the previous simulation. Obviously, these figures are purely indicative, since the actual revenue performance also depends on the distribution of other variables, and in particular the valuations, which were not elicited by our experiment and hence were not calibrated in the above simulations. Nonetheless, as shown in \autoref{sssec:analytic_results}, our analytic results imply that the PPA revenues would always be strictly higher than the PPI for any $\rho>0$, and higher than the PPC for any $\alpha>0$.

\section{Discussion}
\label{sec:discussion}

Our analysis explicitly accounts for the possibility that bidders may value ads beyond the clicks they generate.
The literature's benchmark, in which only clicks are valued, is embedded as a special case in the PPA scheme.
Thanks to this generalisation, our study produces novel theoretical insights on the revenue ranking of PPI and PPC schemes, as well as on the novel auction format we propose in this article.

We have shown that the PPA second-price auction has the same desirable properties (namely, strategy-proofness and efficiency) as its PPI and PPC counterparts. Revenues are identical under the three formats if bidders are fully sophisticated and if attention probabilities are either constant across bidders or uncorrelated with their valuations. But PPA's revenues are higher than the PPI if valuations and attention probabilities are positively correlated,
and they are higher than the PPC as soon as some of the bidders are subject to framing effects.
PPA's revenues could be lower than PPI's only for correlation structures (e.g., with negative correlation between valuations and attention probabilities) under which also the PPC would do worse than the PPI.
Since the PPC is widely considered to outperform the PPI, the possibility of such environments seems less relevant in practice.

To the extent that higher valuation advertisers have stronger incentives to invest in better advertisements,
a positive correlation between valuations and attention probabilities should be expected on average;
i.e. situations in which the PPA outperforms the PPI.
It is also expected that at least a small fraction of bidders are not fully sophisticated;
i.e. situations in which the PPA outperforms the PPC.
If we consider the possibility that different formats affect the incentives of the platform to maximise CTRs and users' attention, there would be even stronger reasons to prefer the PPA over the current alternatives, because it would align the platform's incentives with a more direct measure of the advertisers' objectives.
This would have the effect of increasing the total surplus,
and hence increase revenues beyond the effects covered by our analysis.

The fact that estimated probabilities may be strictly between $0$ and $1$ reflects the uncertainty of the information we may have on whether the ad was actually noticed ($1.0$) or not ($0.0$). However, as long as the estimates are consistent, whenever the probability is estimated at, say, $p=0.7$ it suggests there is a high chance that the ad was truly noticed. Hence, the fact that payment is proportional to the estimated $p$ already ensures that, for large numbers, advertisers are paying the right proportion of times: the lower the probability, the lower the payment, and hence the revenues.
Different payment schemes (e.g., PPI) would generate lower revenues in a different way, i.e. through lower bids, since, under the PPI scheme, bidders would understand that they would be charged also for ads which are ineffective -- the point of the PPA auction is precisely to mitigate this effect, without going all the way to the PPC scheme, in which payments are not made even if the ad is noticed but not clicked.

In this line, perhaps other metrics related to ad performance
could be taken into account in the PPA scheme or derivatives.
For example, the amount of time an ad is shown on screen may increase the chance of noticing the ad.
However, to the best of our knowledge, on-screen time is not factored in the pricing of existing auction formats.
Of course, one could devise an alternative auction format which takes time, rather than attention, as input for the pricing scheme, but we think that it would be a less direct and hence less effective method than the format we propose.
The reason is that the ultimate source of surplus is always the attention of users.
A longer on-screen time may make it more probable that an ad will be eventually noticed,
but it is not obvious that time \emph{per se} may create extra value \emph{independent of}
the increase in the probability of noticing the ad that it may generate.

One of the reasons why the PPC is often preferred to the PPI in practice is because it insures advertisers against the risk of paying for ineffective ads. In the PPI, in contrast, it is the seller who is fully insured, in that there is no uncertainty associated to the payments they receive. The PPA scheme provides an intermediate allocation of risk. Hence, in situations in which both sides exhibit some degree of risk aversion, the PPA may actually increase the total surplus, as well as provide a more equitable split thereof.
A systematic analysis of the impact of risk-aversion on the different payment schemes, however, is left as an opportunity for future work.

The results on demographics discussed at the end of~\autoref{sec:predicting_attention} suggest perceptual differences across the examined user groups. For example, the findings reported in Sections~\ref{sssec:gender} and \ref{sssec:age} suggest differences in the reliability of the attention predictions across gender and age groups, though gender was not found to be a statistically significant confounding factor. Demographics information may be used to fine-tune the design of the PPA auction, so as to further increase its profitability or to pursue other kind of desiderata. More specifically, it could be used in further developments of the PPA auction by re-weighting the way bids affect advertisers' payments, as a function of the observable demographics. We note that, in our analysis, we do not assume to have prior knowledge of any of such a demographic information, nor we use gender and age attributes as part of the training input to the BLSTM model. However, we argue that inferring these attributes is possible even for a commercial Web Search service that does not intentionally store such user profile information~\cite{Hu2007,Pentel2017,7022601}.

At first sight, one might think that the PPA auction will only benefit the advertisers, because it ensures that they are only charged if their ads are actually noticed. However, as shown by our results, it is also beneficial for the platform, since it may boost its revenues.
Ultimately, the PPA auction promotes a fairer and more transparent auction process. The reason is that it directly prices user attention, the PPA provides a more effective target for advertiser campaigns,
thereby further aligning the ad delivery platforms incentives with the objectives of the advertisers.
As mentioned in the \nameref{sec:introduction} section,
this is especially the case for advertisers whose campaigns aim to generate mainly brand or product awareness,
rather than to induce direct online sales, which is the case for many (if not most) of the highest-value
advertisers.

The PPA scheme depends on a diagnostic technology to effectively capture the user attention,
therefore in principle it is more difficult to put into production than the existing PPC and PPI schemes.
In PPC, a simple re-direct through the host site is sufficient to know that the user clicked the ad.
In PPI, the ad delivery platform site knows that a particular ad was displayed while the host site renders the page.
In PPA, the host site must track the mouse movements, and on exiting the page it must inform the ad delivery platform.
However, it is relatively easy to use JavaScript code to track the mouse cursor movements
and send the data unobtrusively in the background.
It can be accomplished, for example, with two lines of code in Google Analytics:
one line to set the \texttt{onmousemove} event listener
and other line to call the \texttt{ga.send()} method.
In addition, previous work has showcased a scalable technology to log mouse movements,
such as transmitting the data whenever a mouse pause is detected~\cite{Huang:2011:NCN:1978942.1979125}
or even using LZW compression to save bandwidth~\cite{Leiva:2013:WBB:2540635.2529996}.
On the other hand, the host site does not need to compute ad attention, nor it should do it.
Instead, computing ad attention should be performed by the ad delivery platform,
not only because some host sites may have limited computational resources,
but also to prevent potential ad fraud.
So, as we see it, the host site only needs to transmit the mouse movements to the ad delivery platform,
and this can be done easily and at scale.
Then, the ad delivery platform must query a trained model, which usually takes a few milliseconds,
to re-estimate ad valuations according to the predicted attention probability.
However, computing ad attention does not need to be done in real-time.
Instead, each ad could be queued for a few seconds before effectively charging the advertiser.

\section{Limitations and Future Work}
\label{limitations}

Our work comes with certain limitations that we intend to address in future work.
First and foremost, the proposed PPA scheme has been introduced for the single-slot auction case,
and it is clear that SERPs display more than one ad at the same time, at different positions.
Now that we know that the PPA auction scheme is feasible,
an important extension we plan to pursue is to investigate the case
in which multiple slots are sold at the same time.
This is challenging from an economic analysis perspective
because of the well-known complexity associated to ensuring strategy-proofness in multi-unit settings (see, e.g., Facebook's VCG mechanism), or for the complexity of strategic behavior in the (non strategy-proof) multi-unit versions of the second-price auctions (such as Google's GSP auction format). Nonetheless, the logic of the PPA pricing scheme can be extended in both directions, and we expect that the advantages discussed above for the single-item case would extend to multi-unit environments as well.

The multi-slot auction case is also challenging for our diagnostic technology
because mouse cursor movements may not clearly indicate user attention to every possible ad on the SERP.
The reason for that is because current SERPs usually include a variable number of modules such as advertisements, query suggestions, video and image results, and media-rich vertical content that compete for users' attention.
There is evidence to support the claim that increasing the number of modules on the SERP may affect task completion~\cite{Rosenholtz05}.
Furthermore, the diversity of modules in SERPs can also impact user experience and scan order~\cite{McCay12, Marcos2015}. One way to circumvent this would be incorporating contextual information about the SERP structure,
at the expense of making our BLSTM model less generalizable.
Note that, currently the input to our BLSTM model is a sequence of raw mouse cursor coordinates only.
In addition to trying to solve these challenges,
we will examine other dependent variables, such as perceived ad relevance or usefulness,
which may be used to inform fine-grained auction schemes.

Second, our experimental methodology, and in particular the way we split our query sample during the cross-validation, may have resulted in some artifacts. More specifically, we introduced multiple instances of the same queries for each combination of ad format and ad position. Ideally, one would like a model to generalize to previously unseen queries. We argue that the way we chose to sample the search queries, i.e. so that they span across several popular and diverse topics, and over a period of 12 months, helped mitigate unaccounted, systematic biases in our analysis. This is further supported by prior evidence which has demonstrated that mouse cursor patterns are, to some extent, independent of the web page content~\cite{Arapakis:2014:UWE:2661829.2661909, Lagun:2014:DCM:2556195.2556265} and, as an extension, of the web search queries as well. Nevertheless, we cannot discount entirely the possibility that our collection of web search queries may have had an effect on the mouse cursor behaviour. Therefore, we plan to investigate this further in follow-up work.

Third, the experimental approach to studying user attention through simulated web search tasks may have affected the ecological validity of results. However, unlike other possible methodologies that could be adopted to analyse web browsing behaviour at large scale (e.g., bucket testing or query log analysis), crowdsourcing allows for exploring a wider range of parameters in a more controlled manner. The downside is, as noted, the difficulty of generalising the findings, because the sample size that can be collected in a crowdsourcing study (e.g., hundreds or thousands of users) is typically much smaller than the sample size that can be collected in the wild (e.g., millions of users).
Ultimately, we decided to ensure to our best effort the internal validity of the experiment;
i.e. the extent to which an observed effect is due to the test conditions. While taking into consideration these limitations, we devised our experimental design so that it mitigates most of the unwanted effects. For example, we introduced brief search tasks that offered a context that the users could relate to, by engaging them with popular web search queries. We further allowed the users to conduct the study from their environment of choice, whether that was e.g., their home or office. Moreover, we did not impose any limitations on the task duration and, last, we collected the mouse cursor data in a non-invasive way that did not disrupt the natural flow of the search task.

We have shown that it is possible to detect user attention to online ads with competitive accuracy.
And, more importantly, we have shown that is possible to do so unobtrusively and at large scale.
Still, there is an opportunity to improve further the prediction capabilities of our diagnostic technology.
For example, by stacking more recurrent layers or increasing the number of neurons,
using other activation functions, improving the optimizer hyperparameters, or implementing more sophisticated model architectures.
For example, the multi-headed self-attention mechanism for RNNs~\cite{Vaswani17}
has shown promise in Natural Language Processing tasks such as Machine Translation and Speech Recognition,
therefore it could be explored for mouse cursor trajectories as well.

Our work has presented some preliminary findings on the potential effects of user's age on attention prediction.
We observed perceptual differences in performance, across different age groups,
that are in line with previous findings from the motor control literature.
These findings may have implications for the proposed diagnostic technology and, by extension, to the PPA auction.
Although in this work we did not explore the role of user's age in the bidding process,
we believe it is definitely worthy of investigation
and thus leave this extension to our auction scheme for follow-up work.
We note that the general idea of the PPA auction scheme, as well as our diagnostic technology, lend themselves to many extensions such as this one.

Another promising venue for future work is the examination of other implicit behavioural signals, such as eye movements, brain activity (which can be measured by means of electroencephalography), and other potentially more objective metrics of attention. While these sensory channels are not as easy to collect as mouse cursor activity, they can nevertheless provide more accurate insights about how users perceive and react to different ad formats. So far, the role of these signals in web search have been studied mostly in isolation~\cite{Goldberg:2002:ETW:507072.507082, Barral:2016:ERA:3022844.3022854, Jacucci19} and we believe that a combination of them would provide us with more insightful information.

Finally, we should mention that our diagnostic technology has been studied in a desktop setting only.
It is not expected to work without adjustments on touch-capable devices such as tablets and smartphones.
This limitation may raise some concerns about the practicality of this diagnostic technology,
since currently half of the web traffic is mobile.
However, it has been reported elsewhere\footnote{https://www.perficientdigital.com/insights/our-research/mobile-vs-desktop-usage-study}$^{,}$\footnote{https://hostingfacts.com/internet-facts-stats/} that engagement is higher on desktop. For example, 58\% of time spent on sites is by desktop users and 42\% of time spent on sites is by mobile users. Similar trends are reported for the percentage of page views per visit. In other words, desktop search is still very relevant and amounts for a profitable and sizeable percentage of web traffic,
and hence it provides a reasonable and important starting point for our analysis.
Potential extensions of our diagnostic technology to account for touch-based interactions include, for example, tracking zoom/pinch gestures and scroll activity instead of the mouse cursor position. This was in fact investigated in previous work by \citet{Guo:2013:MTI:2484028.2484100}, who proposed the Mobile Touch Interaction model, a feature-based classifier that could identify basic patterns of reading and scanning behaviour. Unfortunately, all the proposed touch-based features were found to be weakly correlated with explicit judgements of document relevance. Therefore, there is still plenty of room for improvement in this research area.

\section{Conclusion}
\label{sec:conclusion}

We have introduced the PPA auction scheme, a novel pay-per-attention second-price auction format
that includes user attention to ads in the bidding process.
Under the PPA scheme, advertisers are charged only if their ads are actually noticed by the users.
We have proved that PPA inherits the same desirable properties of the popular PPI and PPC formats (namely, strategy-proofness and efficiency). We also have proved that, in any environment, the revenues of the PPA auction are never lower than its PPI and PPC counterparts, and that in many relevant economic environments they are in fact strictly higher than those of the PPI and PPC formats.

To make PPA feasible, we have introduced a scalable diagnostic technology that estimates user attention to ads in sponsored search using mouse cursor information and a recurrent neural network.
We have tested up to four different classifiers that achieve reasonable performance,
and showed further noticeable improvements by our recurrent neural network, which uses raw mouse cursor data,
over the other baseline models, which rely on ad-hoc and domain-specific features about the SERP.

Further, our numerical simulations exemplify the main insights highlighted by our theoretical findings. They also illustrate the significance of our findings in the context of the distribution of attention probabilities provided by our diagnostic technology. Ultimately, this work extends our current understanding about user engagement with ad displays in web search and paves the way towards more intelligent ad auctions and better user models.

\begin{acks}
We thank the anonymous reviewers for their constructive and valuable feedback.
We also thank Malachy Gavan for valuable research assistance work.
L.\,A. Leiva acknowledges support from the Academy of Finland (BAD project).
\end{acks}

\balance

\bibliographystyle{acmart}
\bibliography{acmart}

\end{document}